\documentclass[sigplan,10pt]{acmart}
\acmSubmissionID{49}

\renewcommand\footnotetextcopyrightpermission[1]{}

\acmYear{2021}\copyrightyear{2021}
\setcopyright{rightsretained}
\acmConference[EuroSys '21]{Sixteenth European Conference on Computer Systems}{April 26--28, 2021}{Online, United Kingdom}
\acmBooktitle{Sixteenth European Conference on Computer Systems (EuroSys '21), April 26--28, 2021, Online, United Kingdom}
\acmPrice{}
\acmDOI{10.1145/3447786.3456228}
\acmISBN{978-1-4503-8334-9/21/04}

\usepackage{enumitem}
\usepackage{booktabs}
\usepackage{soul}
\usepackage{xspace}
\usepackage{color}
\usepackage{xcolor}
\usepackage{upquote}
\usepackage{listings}
\usepackage{amsmath}
\usepackage{wrapfig}
\usepackage{syntax}
\usepackage{caption}
\usepackage{picins}

\usepackage[frozencache]{minted}
\makeatletter
\AtBeginEnvironment{minted}{\dontdofcolorbox}
\def\dontdofcolorbox{\renewcommand\fcolorbox[4][]{##4}}
\makeatother

\usepackage{tikz}
\usetikzlibrary{arrows,automata,shapes.misc,shapes.geometric,positioning}

\newcommand{\eg}{{\em e.g.}, }
\newcommand{\ie}{{\em i.e.}, }
\newcommand{\etc}{{\em etc.}\xspace}
\newcommand{\vs}{{\em vs.} }
 
\newcommand{\heading}[1]{\vspace{4pt}\noindent\textbf{#1}\enspace}
\newcommand{\ttt}[1]{\mintinline[fontsize=\small]{bash}{#1}}
\newcommand{\ttiny}[1]{\mintinline[fontsize=\footnotesize]{bash}{#1}}
\newcommand{\tti}[1]{\texttt{\scriptsize #1}}

\newcommand{\cn}[1]{\mbox{\textcircled{\footnotesize #1}}}
\newcommand{\tcn}[1]{\mbox{\textcircled{\scriptsize #1}}}

\newcommand{\sta}{\cn{\textsc{S}}\xspace}
\newcommand{\pur}{\cn{\textsc{P}}\xspace}
\newcommand{\npu}{\cn{\textsc{N}}\xspace}
\newcommand{\sid}{\cn{\textsc{E}}\xspace}

\newcommand{\tsta}{\tcn{\textsc{S}}\xspace}
\newcommand{\tpur}{\tcn{\textsc{P}}\xspace}
\newcommand{\tnpu}{\tcn{\textsc{N}}\xspace}
\newcommand{\tsid}{\tcn{\textsc{E}}\xspace}

\definecolor{mplblue}{HTML}{1f77b4}
\definecolor{mplorange}{HTML}{ffa500}
\definecolor{mplred}{HTML}{d62728}
\definecolor{mplgreen}{HTML}{008000}
\newcommand{\noeager}{{\color{mplgreen}No Eager}\xspace}
\newcommand{\blockingeager}{{\color{mplorange}Blocking Eager}\xspace}
\newcommand{\nosplit}{{\color{mplred}PaSh w/o Split}\xspace}
\newcommand{\pash}{{\color{mplblue}PaSh}\xspace}

\newcommand{\eat}[1]{}

\newcommand{\kk}[1]{[{\color{magenta}kk: #1}]}

\newcommand{\tr}[1]{} %

\definecolor{editorGray}{rgb}{0.95, 0.95, 0.95}
\definecolor{editorOcher}{rgb}{1, 0.5, 0} %
\definecolor{editorGreen}{rgb}{0, 0.5, 0} %

\definecolor{cdb}{rgb}{0.37, 0.62, 0.63} %

\lstdefinelanguage{sh}{
  morekeywords={for, in, do, done, \|},
  keywordstyle=\color{purple}\ttfamily,
  ndkeywordstyle=\color{black}\ttfamily\bfseries,
  identifierstyle=\color{black},
  sensitive=false,
  comment=[l]{\#},
  commentstyle=\color{lightgray},
  stringstyle=\color{darkgray}\ttfamily,
  morestring=[b]',
  morestring=[b]",
  abovecaptionskip=0pt,
  aboveskip=0pt,
  belowcaptionskip=0pt,
  belowskip=0pt,
  frame=none                     %
}

\usepackage{booktabs}   %
\usepackage{subcaption} %

\title{\sys: Light-touch Data-Parallel Shell Processing}         %

\author{Nikos Vasilakis}
\authornote{The two marked authors contributed equally to the paper.}
\affiliation{MIT}
\email{nikos@vasilak.is}

\author{Konstantinos Kallas}
\authornotemark[1]
\affiliation{University of Pennsylvania}
\email{kallas@seas.upenn.edu}

\author{Konstantinos Mamouras}
\affiliation{Rice University}
\email{mamouras@rice.edu}

\author{Achil\-les Benetopoulos}
\affiliation{Unaffiliated}
\email{abenetopoulos@gmail.com}

\author{Lazar Cvetkovi\'{c}}
\affiliation{University of Belgrade}
\email{cl203023m@student.etf.bg.ac.rs}

\renewcommand{\shortauthors}{N. Vasilakis*, K. Kallas*, K. Mamouras, A.  Benetopoulos, L. Cvetkovi\'{c}}

\newcommand{\cf}[1]{(\emph{Cf}.\S\ref{#1})}
\newcommand{\sx}[1]{(\S\ref{#1})}
\newcommand{\sys}{{\scshape PaSh}\xspace}
\newcommand{\unix}{{\scshape Unix}\xspace}

\setlist{noitemsep,leftmargin=10pt,topsep=2pt,parsep=2pt,partopsep=2pt}

\begin{document}



\begin{abstract}
This paper presents \sys, a system for parallelizing POSIX shell scripts. 
Given a script, \sys converts it to a dataflow graph, performs a series of semantics-preserving program transformations that expose parallelism, and then converts the dataflow graph back into a script---one that adds POSIX constructs to explicitly guide parallelism coupled with \sys-provided \unix-aware runtime primitives for addressing per\-for\-mance- and correctness-related issues.
A lightweight annotation language allows command developers to express key parallelizability properties about their commands.
An accompanying parallelizability study of POSIX and GNU commands---two large and commonly used groups---guides the annotation language and optimized aggregator library that \sys uses.
\sys's extensive evaluation over 44 unmodified \unix scripts shows significant speedups ($0.89$--$61.1\times$, avg: $6.7\times$) stemming from the combination of its program transformations and runtime primitives.
\end{abstract}

\maketitle
\pagestyle{plain}

\section{Introduction}
\label{intro}

The \unix shell is an environment---often interactive---for composing programs written in a plethora of programming languages.
This language-agnosticism, coupled with \unix's toolbox philosophy~\cite{mcilroy1978unix}, makes the shell the primary choice for specifying succinct and simple pipelines for data processing, system orchestration, and other automation tasks.
Unfortunately, parallelizing such pipelines requires significant effort shared between two different programmer groups: %

\begin{itemize}
  \item \emph{Command developers}, responsible for implementing individual commands such as \ttt{sort}, \ttt{uniq}, and \ttt{jq}.
  These developers usually work in a single programming language, leveraging its abstractions to provide parallelism whenever possible.
  As they have no visibility into the command's uses, they expose a plethora of ad-hoc command-specific flags such as \ttt{-t}, \ttt{-}\ttt{-parallel}, \ttt{-p}, and \ttt{-j} %
\cite{pasetto2011comparative, mcilroy1993engineering, stallman1991gnu}.

  \item \emph{Shell users}, who use POSIX shell constructs to combine multiple such commands from many languages into their scripts and are thus left with only a few options for incorporating parallelism.
One option is to use manual tools such as GNU \ttt{parallel}~\cite{Tange2011a}, \ttt{ts}~\cite{tsp}, \ttt{qsub}~\cite{gentzsch2001sun}, \textsc{SLURM}~\cite{yoo2003slurm};
  these tools are either command-unaware, and thus at risk of breaking program semantics, or too coarse-grained, and thus only capable of exploiting parallelism at the level of entire scripts rather than individual components.
Another option is to use shell primitives (such as \ttt{&}, \ttt{wait}) to explicitly induce parallelism, at a cost of manual effort to split inputs, rewrite scripts, and orchestrate execution---an expensive and error-prone process.
To top it off, all these options assume a good understanding of parallelism;
  users with domain of expertise outside computing---from hobbyists to data analysts---are left without options.
\end{itemize}

\begin{figure}[t]
\centering
\includegraphics[width=0.49\textwidth]{\detokenize{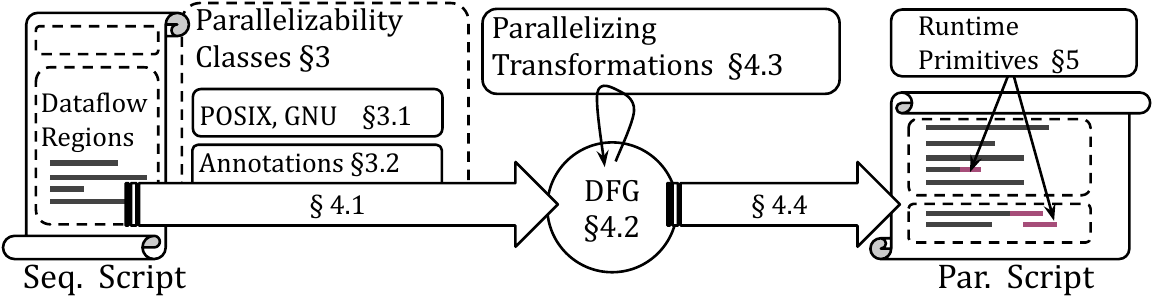}}
\caption{
  \textbf{\sys overview.}
  \sys identifies dataflow regions~\sx{dataflow-regions}, 
  converts them to dataflow graphs~\sx{graph-components},
  applies transformations~\sx{ir:transformations} based on the parallelizability properties  of the commands in these regions (\S\ref{cmd}, \S\ref{ext}),
  and emits a parallel script~\sx{backend} that uses custom primitives~\sx{optimizer}.
}
\vspace{-18pt}
\label{fig:overview}
\end{figure}

\noindent
This paper presents a system called \sys and outlined in Fig.~\ref{fig:overview} for parallelizing POSIX shell scripts that benefits both programmer groups, %
  with emphasis on shell users.
Command developers are given a set of abstractions, akin to lightweight type annotations, for expressing the parallelizability properties of their commands:
  rather than expressing a command's full observable behavior, these annotations focus primarily on its interaction with state.
Shell users, on the other hand, are provided with full automation:
  \sys  analyzes their scripts and extracts latent parallelism.
\sys's transformations are conservative, in that they do not attempt to parallelize fragments that lack sufficient information---\ie at worst, \sys will choose to not improve performance rather than risking breakage.

To address cold-start issues, \sys comes with a library of parallelizability annotations for commands in POSIX and GNU Coreutils.
These large classes of commands serve as the shell's standard library, expected to be used pervasively. %
The study that led to their characterization also informed \sys's annotation and transformation components.

These components are tied together with \sys's runtime component.
Aware of the \unix philosophy and abstractions, it packs a small library of highly-optimized data aggregators as well as high-performance primitives for eager data splitting and merging.
These address many practical challenges %
  and were developed by uncovering several pathological situations, on a few of which we report.

We evaluate \sys on 44 unmodified scripts including 
  (i) a series of smaller scripts, ranging from classic \unix one-liners to modern data-processing pipelines, and (ii) two large and complex use cases for temperature analysis and web indexing.
Speedups range between 0.89--61.1$\times$ (avg: $6.7\times$), with the 39 out of 44 scripts seeing non-trivial speedups. 
\sys's runtime primitives add to the base speedup extracted by \sys's program transformations---\eg 8.83$\times$ over a base 5.93$\times$ average for 10 representative \unix one-liners.
\sys accelerates a large program for temperature analysis by 2.52$\times$, parallelizing both the computation (12.31$\times$) and the preprocessing (2.04$\times$) fragment (\ie data download, extraction, and cleanup), the latter traditionally falling outside of the focus of conventional parallelization systems---even though it takes $75\%$ of the total execution time.

The paper is structured as follows.
It starts by introducing the necessary background on shell scripting and presenting an overview of \sys~\sx{bg}.
Sections \ref{parallelizability}--\ref{impl} highlight key contributions:
\begin{itemize}
  \item
  \S\ref{parallelizability}
    studies the parallelizability of shell commands, and introduces a lightweight annotation language for commands that are executable in a data-parallel manner.

  \item
  \S\ref{ir} presents a dataflow model and associated transformations that expose data parallelism while preserving the semantics of the sequential program.

  \item
  \S\ref{impl} details \sys's runtime component, discussing the correctness and performance challenges it addresses.

\end{itemize}

\noindent
After \sys's evaluation~\sx{eval} and  comparison with related work~\sx{related}, the paper concludes~\sx{discussion}.

\section{Background and Overview}
\label{bg}

This section reviews \unix shell scripting through an example~\sx{bg:pipelines}, later used to explore parallelization challenges~\sx{bg:challenges} and how they are addressed by \sys~\sx{bg:overview}.

\subsection{Running Example: Weather Analysis}
\label{bg:pipelines}

Suppose an environmental scientist    %
  wants to get a quick sense of trends in the maximum temperature across the U.S. over the past five years.
As the National Oceanic and Atmospheric Administration (NOAA) has made historic temperature data publicly available~\cite{noaa}, 
  answering this question is only a matter of a simple data-processing pipeline.

Fig.~\ref{fig:example}'s script starts by pulling the yearly index files and filtering out URLs that are not part of the compressed dataset.
It then downloads and decompresses each file in the remaining set, extracts the values that indicate the temperature, and filters out bogus inputs marked as \ttt{999}.
It then calculates the maximum yearly temperature by sorting the values and picking the top element.
Finally, it matches each maximum value with the appropriate year in order to print the result.
The effort expended writing this script is low:
  its data-processing core amounts to 12 stages and, when expressed as a single line, is only 165 characters long.
This program is no toy:
  a Java program implementing only the last four stages takes 137~LoC~\cite[\S2.1]{hadoop:15}.
To enable such a succinct program composition, \unix incorporates several features.

\heading{\unix Features}
Composition in \unix is primarily achieved with pipes (\ttt{|}), a
construct that allows for task-parallel execution of two commands by
connecting them through a character stream.
This stream is comprised of contiguous character lines separated by newline characters (\textsc{NL}) delineating individual stream elements.
For example, Fig~\ref{fig:example}'s first \ttt{grep} outputs (file-name) elements containing \ttt{gz}, which are then consumed by \ttt{tr}.
A special end-of-file (\textsc{EOF}) condition marks the end of a stream.

Different pipeline stages process data concurrently and possibly at different rates---\eg the second \ttt{curl} produces output at a significantly slower pace than the \ttt{grep} commands before and after it.
The \unix kernel facilitates scheduling, communication, and synchronization behind the scenes.

\begin{figure}[t]
\centering
\begin{minted}[fontsize=\footnotesize]{bash}
base="ftp://ftp.ncdc.noaa.gov/pub/data/noaa";
for y in {2015..2019}; do
 curl $base/$y | grep gz | tr -s" " | cut -d" " -f9 |
 sed "s;^;$base/$y/;" | xargs -n 1 curl -s | gunzip |
 cut -c 89-92 | grep -iv 999 | sort -rn | head -n 1 |
 sed "s/^/Maximum temperature for $y is: /"
done
\end{minted}
\caption{
  \textbf{Calculating maximum temperatures per year.}
  The script downloads daily temperatures recorded across the U.S. for the years 2015--2019 and extracts the maximum for every year.
}
\vspace{-15pt}
\label{fig:example}
\end{figure}

Command flags, used pervasively in \unix, are configuration options that the command's developer has decided to expose to its users to improve the command's general applicability.
For example, by omitting  \ttt{sort}'s \ttt{-r} flag that enables reverse sorting, the user can easily get the minimum temperature.
The shell does not have any visibility into these flags; 
  after it expands special characters such as \ttt{~} and \ttt{*}, it leaves parsing and evaluation entirely up to individual commands.

Finally, \unix provides an environment for composing commands written in any language.
Many of these commands come with the system---\eg ones defined by the POSIX standard or ones part of the GNU Coreutils---whereas others are available as add-ons.
The fact that commands are developed in a variety of languages---including shell scripts---provides users with significant flexibility.
For example, one could replace \ttt{sort} and \ttt{head} with \ttt{./avg.py} to get the average rather than the maximum---the pipeline still works, as long as \ttt{./avg.py} conforms to the interface outlined earlier.

\subsection{Parallelization Challenges}
\label{bg:challenges}

While these features aid development-effort economy through powerful program composition, they complicate shell script parallelization, which even for simple scripts such as the one in Fig.~\ref{fig:example} create
several challenges.

\heading{Commands} %
In contrast to restricted programming frameworks that enable parallelization by supporting a few care\-ful\-ly-designed primitives~\cite{streamit:02, brook:04, mapreduce:08, spark:12}, the \unix shell provides an unprecedented number and variety of composable commands.
To be parallelized, each command may require special analysis and treatment---\eg exposing data parallelism in Fig.~\ref{fig:example}'s \ttt{tr} or \ttt{sort} would require splitting their inputs, running them on each partial input, and then merging the partial results.\footnote{
  As explained earlier~\sx{intro}, commands such as \ttiny{sort} may have \emph{ad hoc} flags such as \ttiny{--parallel}, which do not compose across commands and may risk breaking correctness or not exploiting performance potential~\sx{micro}.
}
Automating such an analysis is infeasible, as individual commands are black boxes written in a variety of programming languages and models.
Manual analysis is also challenging, due to the sheer number of commands and the many flags that affect their behavior---\eg Fig.~\ref{fig:example}'s program invokes \ttt{cut} with two separate sets of flags.

\heading{Scripts} %
Another challenge is due to the language of the POSIX shell. %
First, the language contains constructs that enforce sequential execution:
   The sequential composition operator (\ttt{;}) in Fig.~\ref{fig:example} indicates that the assignment to \ttt{base} must be completed before everything else.
Moreover, the language semantics only exposes limited task-based parallelism in the form of constructs such as \ttt{&}. %
Even though Fig.~\ref{fig:example}'s \ttt{for} focuses only on five years of data, \ttt{curl} still outputs thousands of lines per year;
  naive parallelization of each loop iteration will miss such opportunities.
Any attempt to automate parallelization should be aware of the POSIX shell language, exposing latent data parallelism without modifying execution semantics.

\heading{Implementation} %
On top of command and shell semantics, the broader \unix environment has its own set of quirks.
Any attempt to orchestrate parallel execution will hit challenges related to task parallelism, deadlock prevention, and runtime performance.
For example, forked processes piping their combined results to Fig.~\ref{fig:example}'s \ttt{head} may not receive a \ttt{PIPE} signal if \ttt{head} exits prior to opening all pipes.
Moreover, several commands such as \ttt{sort} and \ttt{uniq} require specialized data aggregators in order to be correctly parallelized.

\subsection{\sys Design Overview}
\label{bg:overview}

At a high level, \sys takes as input a POSIX shell script like the one in Fig.~\ref{fig:example} and outputs a new POSIX script that incorporates data parallelism.
The degree of data parallelism sought by \sys is configurable using a \ttt{--width} parameter, whose default value is system-specific.
Fig.~\ref{fig:example2} highlights a few fragments of the parallel script resulting from applying \sys with \ttt{--width=2} to the script of Fig.~\ref{fig:example}---resulting in 2 copies of \{\ttt{grep}, \ttt{tr}, \ttt{cut}, {\em etc.}\}.

\sys first identifies sections of the script that are potentially parallelizable, \ie lack synchronization and scheduling constraints, and converts them to dataflow graphs (DFGs).
It then performs a series of DFG transformations that expose parallelism without breaking semantics, by expanding the DFG to the desired \ttt{width}.
Finally, \sys converts these DFGs back to a shell script augmented with \sys-provided commands. %
The script is handed off to the user's original shell interpreter for execution.
\sys addresses the aforementioned challenges~\sx{bg:challenges} as below.

\heading{Commands}
To understand standard commands available in any shell, \sys groups POSIX and GNU commands into a small but well-defined set of \emph{parallelizability classes}~\sx{cmd}.
Rather than describing a command's full observable behavior, these classes focus on information that is important for data parallelism.
To allow other commands to use its transformations, \sys defines a light \emph{annotation language} for describing a command's parallelizability class~\sx{ext}.
Annotations are expressed once per command rather than once per script and are aimed towards command developers rather than its users, so that they can quickly and easily capture the characteristics of the commands they develop.

\begin{figure}[t]
\centering
\begin{minted}[fontsize=\footnotesize,escapeinside=()]{bash}
mkfifo $t{0,1...}
curl $base/$y > $t0 & cat $t0 | split $t1 $t2 &
cat $t1 | grep gz > $t3 & 
cat $t2 | grep gz > $t4 &
...
cat $t9 | sort -rn > $t11 & cat $t10 | sort -rn > $t12 &
cat $t11 | eager > $t13 & cat $t12 | eager > $t14 &
sort -mrn $t13 $t14 > $t15 &
cat $t15 | head -n1 > $out1 &
wait $! && get-pids | xargs -n 1 kill -SIGPIPE
\end{minted}
\caption{
  \textbf{Output of \ttiny{pash --width=2} for Fig.~\ref{fig:example} (fragment).}
  \sys orchestrates the parallel execution through named pipes, parallel operators, and custom runtime primitives---\eg \ttiny{eager}, \ttiny{split}, and \ttiny{get-pids}.
}
\vspace{-15pt}
\label{fig:example2}
\end{figure}

\heading{Scripts}
To maintain sequential semantics, \sys first analyzes a script to identify \emph{dataflow regions} containing commands that are candidates for parallelization~\sx{dataflow-regions}.
This analysis is guided by the script structure: %
  some constructs expose parallelism (\eg \ttt{&}, \ttt{|}); others enforce synchronization (\eg \ttt{;}, \ttt{||}).
\sys then converts each dataflow region to a \emph{dataflow graph} (DFG)~\sx{graph-components}, 
  a flexible representation that enables a series of local transformations to expose data parallelism, converting the graph into its parallel equivalent~\sx{ir:transformations}.
Further transformations compile the DFG back to a shell script that uses POSIX constructs to guide parallelism explicitly while aiming at preserving the semantics of the sequential program~\sx{backend}.

\heading{Implementation}
\sys addresses several practical challenges through a set of constructs it provides---\ie 
modular components for augmenting command composition~\sx{impl}.
It also provides a small and efficient \emph{aggregator library} targeting a large set of parallelizable commands.
All these commands live in the \ttt{PATH} and are addressable by name, which means they can be used like (and by) any other commands.

\section{Parallelizability Classes}
\label{parallelizability}

\sys aims at parallelizing data-parallel commands, \ie ones that can process their input in parallel, encoding their characteristics by assigning them to
\emph{parallelizability classes}.
\sys leans towards having a few coarse classes rather than many detailed ones---among other reasons, to simplify their understanding and use by command developers.

This section starts by defining these classes, along with a parallelizability study of the commands in {\sc POSIX} and GNU Coreutils~\sx{cmd}.
Building on this study, it develops a lightweight extensibility framework that enables light-touch parallelization of a command by its developers~\sx{ext}.
\sys in turn uses this language to annotate POSIX and GNU commands and generate their wrappers, as presented in later sections.

\subsection{Parallelizability of Standard Libraries}
\label{cmd}

Broadly speaking, shell commands can be split into four major classes with respect to their parallelization characteristics, depending on what kind of state they mutate when processing their input (Tab.\ref{tab:classes}).
These classes are ordered in ascending difficulty (or impossibility) of parallelization.
In this order, some classes can be thought of as subsets of the next---\eg all stateless commands are pure---meaning that the synchronization mechanisms required for any superclass would work with its subclass (but foregoing any performance improvements).
Commands can change classes depending on their flags,
which are discussed later~\sx{ext}.

\begin{table}[t]
\center
\footnotesize
\setlength\tabcolsep{3pt}
\caption{
  \footnotesize{
    \textbf{Parallelizability Classes}.
    Broadly, \unix commands can be grouped into four classes according to their parallelizability properties.
  }
}
\begin{tabular*}{\columnwidth}{l @{\extracolsep{\fill}} llll}
\toprule
Class                    &  Key    & Examples                                    & Coreutils              & POSIX       \\
\midrule
Stateless                & ~\tsta  & \tti{tr},   \tti{cat},    \tti{grep}        &  13 (12.5\%)           & 19 (12.7\%)        \\  %
Parallelizable Pure      & ~\tpur  & \tti{sort}, \tti{wc},     \tti{head}        &  17 (16.3\%)           & 13 (8.7\%)          \\  %
Non-parallelizable Pure  & ~\tnpu  & \tti{sha1sum}                               &  13 (12.5\%)           & 11 (7.3\%)       \\  %
Side-effectful           & ~\tsid  & \tti{env},  \tti{cp}, \tti{whoami}          &  61 (58.6\%)           & 105 (70.4\%)          \\  %
\bottomrule
\vspace{-18pt}
\end{tabular*}
\label{tab:classes}
\end{table}

\heading{Stateless Commands}
The first class, \sta, contains commands that operate on individual line elements of their input, without maintaining state across invocations.
These are commands that can be expressed as a purely functional \emph{map} or \emph{filter}---\eg \ttt{grep} filters out individual lines and \ttt{basename} removes a path prefix from a string.
They may produce multiple elements---\eg \ttt{tr} may insert {\sc NL} tokens---but always return empty output for empty input.
Workloads that use only stateless commands are trivial to parallelize:
  they do not require any synchronization to maintain correctness, nor caution about where to split inputs.

The choice of line as the data element strikes a convenient balance between coarse-grained (files) and fine-grained (characters) separation while staying aligned with \unix's core abstractions.
This choice can affect the allocation of commands in \sta, as many of its commands (about 1/3) are stateless \emph{within} a stream element---\eg\ttt{tr} transliterates characters within a line, one at a time---enabling further parallelization by splitting individual lines.
This feature may seem of limited use, as these commands are computationally inexpensive, precisely due to their narrow focus.
However, it may be  useful for cases with very large stream elements (\ie long lines) such as the \ttt{.fastq} format used in bioinformatics.

\heading{Parallelizable Pure Commands}
The second class, \pur, contains commands that respect functional purity---\ie same outputs for same inputs---but maintain internal state across their entire pass.
The details of this state and its propagation during element processing affect their parallelizability characteristics.
Some commands are easy to parallelize, because they maintain trivial state and are commutative---\eg \ttt{wc} simply maintains a counter.
Other commands, such as \ttt{sort}, maintain more complex invariants that have to be taken into account when merging partial results.

Often these commands do not operate in an online fashion, but need to block until the end of a stream.
A typical example of this is \ttt{sort}, which cannot start emitting results before the last input element has been consumed.
Such constraints affect task parallelism, but not data parallelism:
  \ttt{sort} can be parallelized significantly using divide-and-conquer techniques---\ie by encoding it as a group of (parallel) $map$ functions followed by an $aggregate$ that merges the results.

\heading{Non-parallelizable Pure Commands}
The third class, \npu, contains commands that, while purely functional, cannot be parallelized within a single data stream.\footnote{
  Note that these commands may still be parallelizable across different data streams, for example when applied to different input files.
}
This is because their internal state depends on prior state in non-trivial ways over the same pass. %
For example, hashing commands such as \ttt{sha1sum} maintain complex state that has to be updated sequentially.
If parallelized on a single input, each stage would need to wait on the results of all previous stages, foregoing any parallelism benefits.

It is worth noting that while these commands are not parallelizable at the granularity of a single input, they are still parallelizable across different inputs.
For example, a web crawler involving hashing to compare individual pages would allow \ttt{sha1sum} to proceed in parallel for different pages.

\heading{Side-effectful Commands}
The last class, \sid, contains commands that have side-effects across the system---for example, updating environment variables, interacting with the filesystem, and accessing the network.
Such commands are not parallelizable without finer-grained concurrency control mechanisms that can detect side-effects across the system.

This is the largest class, for two main reasons.
First, it includes commands related to the file-system---a central abstraction of the \unix design and philosophy~\cite{unix}.
In fact, \unix uses the file-system as a proxy to several file-unrelated operations such as access control and device driving. %
Second, this class contains commands that do not consume input or do not produce output---and thus are not amenable to data parallelism.
For example, \ttt{date}, \ttt{uname}, and \ttt{finger} are all commands interfacing with kernel- or hardware-generated information and do not consume any input from user programs.

\subsection{Extensibility Framework}
\label{ext}

To address the challenge of a language-agnostic environment~\sx{bg:challenges}, \sys allows communicating key details about their parallelizability through a lightweight extensibility framework comprising two components:
  an annotation language, and 
  an interface for developing parallel command aggregators.
The framework can be used both by developers of new commands %
  as well as developers maintaining existing commands.
The latter group can express additions or changes to the command's implementation or interface, which is important as commands are maintained or extended over long periods of time.

The extensibility framework is expected to be used by individuals who understand the commands and their parallelizability properties, and thus \sys assumes their correctness.
The framework could be used as a foundation for crowdsourcing the annotation effort, for testing annotation records, and for generating command aggregators.
We use this extension framework in a separate work to synthesize command aggregators automatically~\cite{kumquat}.

\heading{Key Concerns}
\sys's annotations focus on three crucial concerns about a command:
  (C1) its parallelizability class,
  (C2) its inputs and outputs, and the characteristics of its input consumption, and
  (C3) how flags affect its class, inputs, and outputs.
The first concern was discussed extensively in the previous section; we now turn to the latter two.

Manipulating a shell script in its original form to expose parallelism is challenging as each command has a different interface.
Some commands read from standard input, while others read from input files.
Ordering here is important, as a command may read several inputs in a predefined input order.
For example, \ttt{grep "foo" f1 - f2} first reads from \ttt{f1}, then shifts to its standard input, and finally reads \ttt{f2}.

Additionally, commands expose flags or options for allowing users to control their execution. 
Such flags may directly affect a command's parallelizability classification as well as the order in which it reads its inputs.
For example, \ttt{cat} defaults to \sta, but with \ttt{-n} it jumps into \pur because it has to keep track of a counter and print it along with each line.

To address all these concerns, \sys introduces an annotation language encoding first-order logic predicates.
The language allows specifying the aforementioned information, \ie correspondence of arguments to inputs and outputs and the effects of flags.
Annotations assign one of the four parallelizability class as a default class, subsequently refined by the set of flags the command exposes.
Additionally, for commands in \sta and \pur, the language captures how a command's arguments, standard input, and standard output correspond to its inputs and outputs.
Annotations in these classes can also express ordering information about these inputs---effectively lifting commands into a more convenient representation where they only communicate with their environment through a list of input and output files.

The complete annotation language currently contains 8 operators, one of which supports regular expressions.
It was used to annotate 47 commands, totaling 708 lines of JSON---an effort that took about 3--4 hours.
Annotation records are by default conservative so as to not jeopardize correctness, but can be incrementally refined to capture parallelizability when using increasingly complex combinations of flags.
The language is extensible with more operators (as long as the developer defines their semantics);
  it also supports writing arbitrary Python code for commands whose properties are difficult to capture---\eg higher-order \ttt{xargs}, whose parallelizability class depends on the class of the first-order command that it invokes.

\heading{Example Annotations}
Two commands whose annotations sit at opposing ends of the complexity spectrum are \ttt{chmod} and \ttt{cut}.
The fragment below shows the annotation for \ttt{chmod}.
\begin{minted}[fontsize=\footnotesize]{json}
{ "command": "chmod",
  "cases": [ { "predicate": "default", 
               "class": "side-effectful" } ] }
\end{minted}
This annotation is simple, but serves as an illustration of the annotation structure. Each annotation is a JSON record that contains the command name, and a sequence of cases.
Each case contains a predicate that matches on the arguments of the command invocation.
It assigns a parallelizability class (C1) to a specific command instance, \ie the combination of its inputs-output consumption (C2) and its invocation arguments (C3).
In this case, \ttt{chmod} is side-effectful, and thus the \ttt{"default"} predicate of its single \ttt{cases} value always matches---indicating the presence of side-effects.

\tr{\kk{We don't need to designate the stderror, because we assume that it
  is never the main output of a command and it will never be used by
  the input of another command in the pipeline. If someone indeed
  wants to do this, they can just use a redirect I think to get around
  it.}}

\tr{\kk{If we talk about the language it would be good to give some
  statistics about how many commands (out of the ones that we have
  classified) can be represented with one, two clauses etc. If we have
  a different solution (like Python function that given a command and
  its arguments returns the category) then we could talk about how
  many lines of code it took to categorize all the commands that we
  did.}}

\tr{\kk{Note: Since a categorization language will probably not be
  complete (especially the one showing the input arguments) there
  should be a backup mechanism, when the language is not expressive
  enough, like a Python function that categorizes the command, and
  returns its input argument if it is stateless.}}

\tr{\kk{(Maybe) We should have a crisp point about why we have this categorization
  language, and why don't we just allow someone to write a function in
  Python for each command, that given the command and its arguments,
  returns whether the command category. Possible arguments include,
  the fact that this language is simpler to use, especially by non
  experts that just run some script but have installed some commands
  that are not ``supported''. Another possible argument is that we
  could be able to reason about the constructs of the categorization
  lagnauge, and that they will be easier to read and
  understand. Another argument is that since these categorizations
  should be shareable among users, it would be bad to execute
  arbitrary Python code, so using this language, expressivity is
  limited. All of these arguments are a little bit weak though. Niko,
  what do you think?}}

The annotation for \ttt{cut} is significantly more complex, and is only shown in part (the full annotation is in Appendix~\ref{annotation-apx}).
This annotation has two cases, each of which consists of a predicate on \ttt{cut}'s arguments and an assignment of its parallelizability class, inputs, and outputs as described above.
We only show \ttt{cut}'s first predicate, slightly simplified for clarity.
\begin{minted}[fontsize=\footnotesize]{json}
{ "predicate": {"operator": "exists", "operands": [ "-z" ]},
  "class": "n-pure",
  "inputs": [ "args[:]" ],
  "outputs": [ "stdout" ] }
\end{minted}
This predicate indicates that if \ttt{cut} is called with \ttt{-z} as an argument, then it is in \npu, \ie it only interacts with the environment by writing to a file (its \ttt{stdout}) but cannot be parallelized. This is because \ttt{-z} forces \ttt{cut} to delimit lines with \ttt{NUL} instead of newline, meaning that we cannot parallelize it by splitting its input in the line boundaries.
The case also indicates that \ttt{cut} reads its inputs from its non-option arguments.

Experienced readers will notice that \ttt{cut} reads its input from its \ttt{stdin} if no file argument is present.
This is expressed in the \ttt{"options"} part of \ttt{cut}'s annotation, shown below:
\begin{minted}[fontsize=\footnotesize]{json}
{ "command": "cut",
  "cases": [ ... ],
  "options": [ "empty-args-stdin",
               "stdin-hyphen" ] }
\end{minted}
Option \ttt{"empty-args-stdin"} indicates that if non-option arguments are empty, then the command reads from its \ttt{stdin}.
Furthermore, option \ttt{"stdin-hyphen"} indicates that a non-option argument that is just a dash \ttt{-} represents the stdin.

The complete annotation in Appendix~\ref{annotation-apx}) shows the rest of the cases (including the default case for \ttt{cut}, which indicates that it is in \sta).

\heading{Custom Aggregators}
For commands in \sta, the annotations are enough to enable parallelization:
  commands are applied to parts of their input in parallel, and their outputs are simply concatenated.

To support the parallelization of arbitrary commands in \pur, \sys allows supplying custom \emph{map} and \emph{aggregate} functions.
In line with the \unix philosophy, these functions can be written in any language as long as they conform to a few invariants:
  (i) \emph{map} is in \sta and \emph{aggregate} is in \pur,
  (ii) \emph{map} can consume (or extend) the output of the original command and  \emph{aggregate} can consume (and combine) the results of multiple \emph{map} invocations, and
  (iii) their composition produces the same output as the original command.
\sys can use the \emph{map} and \emph{aggregate} functions in its graph transformations~\sx{ir} to further expose parallelism.

Most commands only need an \emph{aggregate} function, as the \emph{map} function for many commands is the sequential command itself.
\sys defines a set of aggregators for many POSIX and GNU commands in \pur.
This set doubles as both \sys's standard library and an exemplar for community efforts tackling other commands.
Below is the Python code for one of the simplest \emph{aggregate} functions, the one for \ttt{wc}:
 
\begin{minted}[fontsize=\footnotesize]{python}
#!/usr/bin/python
import sys, os, functools, utils

def parseLine(s):
  return map(int, s.split())

def emitLine(t):
  f = lambda e: str(e).rjust(utils.PAD_LEN, ' ')
  return [" ".join(map(f, t))]

def agg(a, b):
  # print(a, b)
  if not a:
    return b
  az = parseLine(a[0])
  bz = parseLine(b[0])
  return emitLine([ (i+j) for (i,j) in zip(az, bz) ])

utils.help()
res = functools.reduce(agg, utils.read_all(), [])
utils.out("".join(res))
\end{minted}

\noindent
The core of the aggregator, function \ttt{agg}, takes two input streams as its arguments.
The \ttt{reduce} function lifts the aggregator to arity $n$ to support an arbitrary number of parallel $map$ commands.
This lifting allows developers to think of aggregators in terms of two inputs, but generalize them to operate on many inputs.
Utility functions such as \ttt{read} and \ttt{help}, common across \sys's aggregator library, deal with error handling when reading multiple file descriptors, and offer a \ttt{-h} invocation flag that demonstrates the use of each aggregator.

\sys's library currently contains over 20 aggregators, many of which are usable by more than one command or flag.
For example, the aggregator shown above is shared among \ttt{wc}, \ttt{wc -lw}, \ttt{wc -lm}, \etc

\tr{\kk{I am not sure a general interface is so easy to design. It needs
  more though. It might be beneficial to just talk about sort and wc
  here and how we implemented them and nothing more. Or maybe this
  could then go to the implementation? Or maybe say that one can write
  a Python function that given a node of the graph, transforms it into
  many. I am not sure what is best...}}

\tr{Can we find a solution for the commands in coreutils?}

\section{Dataflow Graph Model}
\label{ir}

\sys's core is an abstract dataflow graph (DFG) model~\sx{graph-components} used as the intermediate representation on which \sys performs parallelism-exposing transformations.
\sys first lifts sections of the input script to the DFG representation~\sx{dataflow-regions}, 
  then performs transformations to expose parallelism (up to the desired \ttt{--width})~\sx{ir:transformations}, 
  and finally instantiates each DFG back to a parallel shell script~\sx{backend}.
A fundamental characteristic of \sys's DFG is that it encodes the
order in which a node reads its input streams (not just the order of input elements per stream), which in turn enables a set of
graph transformations that can be iteratively applied to expose
parallelization opportunities for \tsta and \tpur commands.

To the extent possible, this section is kept informal and intuitive.
The full formalization of the dataflow model, the shell$\leftrightarrow$DFG bidirectional translations, and the parallelizing transformations, as well as their proof of correctness with respect to the script's sequential output, are all presented in a separate work~\cite{handa2020order}.

\subsection{Frontend: From a Sequential Script to DFGs}
\label{dataflow-regions}

\begin{figure}[t]
  \centering
  \includegraphics[width=\columnwidth]{\detokenize{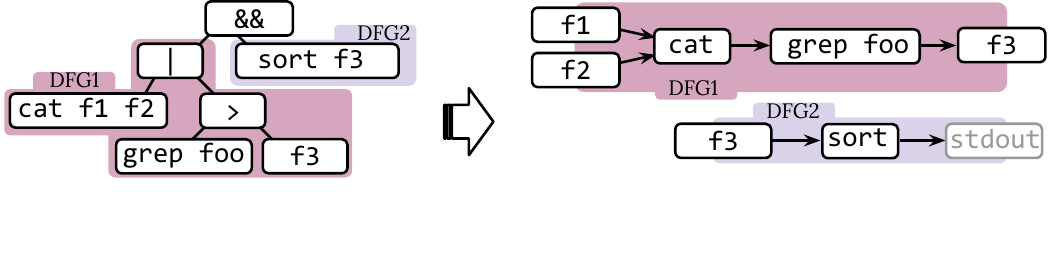}}
  \vspace{-25pt}
  \caption{
    \textbf{From a script AST to DFGs.}
    The AST on the left has two dataflow regions, each not extending beyond \ttiny{&&}~\cf{dataflow-regions}.
		Identifiers \ttt{f1}, \ttt{f2}, and \ttt{f3} sit at the boundary of the DFG.
  }
  \vspace{-15pt}
  \label{fig:ast-dfg}
\end{figure}

\heading{Dataflow Regions}
In order to apply the graph transformations that expose data parallelism, \sys first has to identify program sub-expressions that can be safely transformed to a dataflow graph, 
  \ie sub-expressions that (i) do not impose any scheduling or synchronization constraints (\eg by using \ttt{;}), and (ii) take a set of files as inputs and produce a set of files as outputs.
The search for these regions is guided by the shell language and the structure of a particular program.
These contain information about (i) fragments that can be executed independently, and (ii) barriers that are natural synchronization points. %
Consider this code fragment (Fig.~\ref{fig:ast-dfg}):
\begin{minted}[fontsize=\footnotesize]{bash}
     cat f1 f2 | grep "foo" > f3 && sort f3 
\end{minted}
\noindent
The \ttt{cat} and \ttt{grep} commands execute independently (and concurrently)
in the standard shell, but \ttt{sort} waits for their completion prior to start.
Intuitively, dataflow regions correspond to sub-expressions of
the program that would be allowed to execute independently by
different processes in the POSIX standard~\cite{posix}. Larger
dataflow regions can be composed from smaller ones using the
pipe operator (\ttt{|}) and the parallel-composition operator (\ttt{&}).
Conversely, all other operators, including sequential composition (\ttt{;}) and logical operators (\ttt{&&}, \ttt{||}), represent
barrier constructs that do not allow dataflow regions to expand beyond them.

\heading{Translation Pass}
\sys's front-end performs a depth-first search on the AST of
the given shell program.  During this pass, it extends the
dataflow regions bottom-up, translating their
independent components to DFG nodes until a barrier
construct is reached. All AST subtrees not translatable to DFGs are kept as they are. The output of the translation pass is the
original AST where dataflow regions have been replaced with
DFGs. 

To identify opportunities for parallelization, the translation pass extracts each command's  parallelizability class together with its inputs and outputs.
To achieve this for each command, it searches all its available annotations~\sx{ext} and resorts to conservative defaults if none is found.
If the command is in \sta, \pur, or \npu, the translation pass initiates a dataflow region that is propagated up the tree.

Due to the highly dynamic nature of the shell, some information is not known to \sys at translation time.
Examples of such information include the values of environment variables, unexpanded strings, and sub-shell constructs.
For the sake of correctness, \sys takes a conservative approach and avoids parallelizing nodes for which it has incomplete information.
It will not attempt to parallelize sub-expressions for which the translation pass cannot infer that, \eg  an environment variable passed as an argument to a command does not change its parallelizability class.

\subsection{Dataflow Model Definitions}
\label{graph-components}

The two main shell abstractions are
  (i) data streams, \ie files or pipes, and
  (ii) commands, communicating through these streams.

\heading{Edges---Streams}
Edges in the DFG represent streams, the basic data abstraction of the shell.
They are used as communication channels between nodes in the graph, and as the input or output of the entire graph.
For example, the edges in DFG1 of Figure~\ref{fig:ast-dfg} are the files \ttt{f1}, \ttt{f2}, and \ttt{f3}, as well as the unnamed pipe that connects \ttt{cat} and \ttt{grep}.
We fix the data quantum to be character lines, \ie sequences of characters followed by the newline character,\footnote{
  This is a choice that is not baked into \sys's DFG model, which supports arbitrary data elements such as characters and words, but was made to simplify alignment with many \unix commands.
}
  so edges represent possibly unbounded sequences of lines.
As seen above, an edge can either refer to a named file, an ephemeral pipe, or a \unix FIFO used for interprocess communication.
Edges that do not start from a node in the graph represent the graph inputs;
  edges that do not point to a node in the graph represent its outputs.

\heading{Nodes---Commands}
A node of the graph represents a relation (to capture nondeterminism) from a possibly empty list of input streams to a list of output streams.
This representation captures all the commands in the classes \tsta, \tpur, and \tnpu, since they only interact with the environment by reading and writing to streams.
We require that nodes are monotone, namely that they cannot retract output once they have produced it.
As an example, \ttt{cat}, \ttt{grep}, and \ttt{sort} are the nodes in the DFGs of Figure~\ref{fig:ast-dfg}.

\heading{Streaming Commands}
A large subset of the parallelizable \tsta and \tpur classes falls into the special category of streaming commands. 
These commands have two execution phases.
First, they consume a (possibly empty) set of input streams that act as configuration.
Then, they transition to the second phase where they consume the rest of their inputs sequentially, one element at a time, in the order dictated by the configuration phase and produce a single output stream.
The simplest example of a streaming command is \ttt{cat}, which has an empty first phase and then consumes its inputs in order, producing their concatenation as output. 
A more interesting example is \ttt{grep} invoked with \ttt{-f patterns.txt} as arguments;
  it first reads \ttt{patterns.txt} as its configuration input, identifying the patterns for which to search on its input,
  and then reads a line at a time from its standard input, stopping when it reaches EOF.

\subsection{Graph Transformations}
\label{ir:transformations}

\begin{figure}[t]
\centering
\includegraphics[width=\columnwidth]{\detokenize{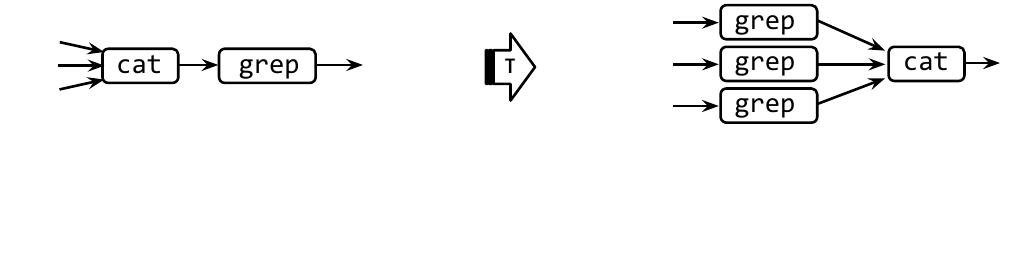}}
\vspace{-45pt}
\caption{
  \textbf{Stateless parallelization transformation.}
  The \ttiny{cat} node is commuted with the stateless node to exploit available data parallelism.
}
\vspace{-18pt}
\label{fig:parallelization-transformation}
\end{figure}

\sys defines a set of semantics-preserving graph transformations
that act as parallelism-exposing optimizations. Both the domain
and range of these transformations is a graph in \sys's DFG model;
  transformations can be composed arbitrarily and in any order.
Before describing the different types of transformations, we formalize
the intuition behind classes \sta and \pur described informally
earlier~\sx{cmd}.

\heading{Stateless and Parallelizable Pure Commands}
Stateless commands such as \ttt{tr} operate independently on individual lines of their input stream without maintaining any state~\sx{cmd}. 
To avoid referring to the internal command state, we can instead determine that a command is stateless if its output is the same if we ``restart'' it after it has read an arbitrary prefix of its input.
If a command was stateful, then it would not produce the same output after the restart.
Formally, a streaming command $f$ is stateless if it commutes with
the operation of concatenation on its streaming input, \ie it is a semigroup homomorphism:
\[
\forall x, x', c, f(x \cdot x', c) = f(x, c) \cdot f(x', c)
\]
\noindent
In the above $x \cdot x'$ is the concatenation of the two parts of $f$'s streaming input and $c$ is the configuration input (which needs to be passed to both instances of $f$).
The above equation means that applying the command $f$ to a concatenation of two inputs $x, x'$ produces the same output as applying $f$ to each input $x, x'$ separately, and concatenating the outputs.
Note that we only focus on deterministic stateless commands and that is why $f$ is a function and not a relation in the above.

Pure commands such as \ttt{sort} and \ttt{wc} can also be parallelized, %
  using divide-and-conquer parallelism.
These commands can be applied independently on different segments of their inputs, and then their outputs are aggregated to produce the final result. 
More formally, these pure commands $f$ can be implemented as a combination of a function $map$ and an associative function $aggregate$ that satisfy the following equation:
\[
\forall x, x', c, f(x \cdot x', c) = aggregate(map(x, c), map(x', c), c)
\]

\heading{Parallelization Transformations}
Based on these equations, we can define a parallelization
  transformation on a node $f \in$ \sta whose streaming input is a concatenation, \ie produced using the command \ttt{cat}, of $n$ input streams and is followed by a node $f'$
(Fig.~\ref{fig:parallelization-transformation}).
The transformation replaces $f$ with $n$ new nodes, routing each of the $n$ input streams
to one of them, and commutes the \ttt{cat} node after them to
concatenate their outputs and transfer them to $f'$. Formally:
\[
  v(x_1 \cdot x_2 \cdots x_n, s) \Rightarrow v(x_1, s) \cdot v(x_2, s) \cdots v(x_n, s)
\]
The transformation can be extended straightforwardly to nodes $v \in$ \pur,
implemented by a $(map, aggregate)$ pair:
\begin{align*}
  v(x_1 & \cdot x_2 \cdots x_n, s) \Rightarrow\\
  & aggregate(map(x_1, s), map(x_2, s), \ldots map(x_n, s), s)
\end{align*}
Both transformations can be shown to preserve the behavior of the original graph assuming that the pair $(map, aggregate)$ meets the three invariants outlined earlier~\sx{ext} and the aforementioned equations hold.

\begin{figure}[t]
\centering
\includegraphics[width=\columnwidth]{\detokenize{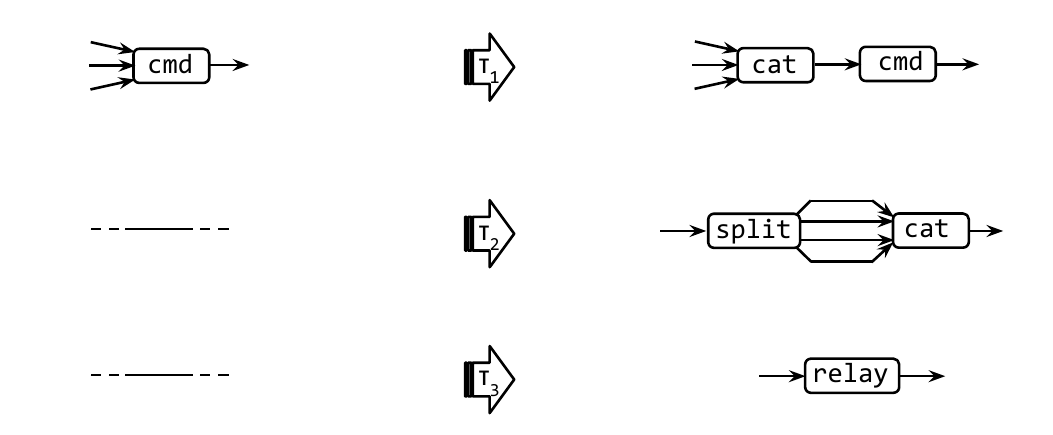}}
\vspace{-20pt}
\caption{
  \textbf{Auxiliary transformations.}
  These augment the DFG with \ttiny{cat}, \ttiny{split}, and \ttiny{relay} nodes.
}
\vspace{-12pt}
\label{fig:auxiliary-transformations}
\end{figure}

\heading{Auxiliary Transformations}
\sys also performs a set of auxiliary transformations $t_{1-3}$ that are
depicted in Fig.~\ref{fig:auxiliary-transformations}. If a node has many
inputs, $t_1$ concatenates these inputs by inserting a \ttt{cat} node
to enable the parallelization transformations. In cases where a
parallelizable node has one input and is not preceded by a
concatenation, $t_2$ inserts a \ttt{cat} node that is preceded by its
inverse \ttt{split}, so that the concatenation can be commuted with
the node. Transformation $t_3$ inserts a relay node that
performs the identity transformation. 
Relay nodes can be useful for performance improvements~\sx{optimizer}, as well as for monitoring and debugging.

\heading{Degree of Parallelism}
The degree of parallelism achieved by \sys is affected by the width of the final dataflow graph.
The dataflow width corresponds, intuitively, to the number of data-parallel copies of each node of the sequential graph and thus the fanout of the \ttt{split} nodes that \sys introduces.
The dataflow width is configured using the \ttt{--width} parameter, which can be chosen by the user depending on their script characteristics, input data, and target execution environment.
By default, \sys assigns width to 2 if it is executing on a machine with 2-16 processors, and \ttt{floor(cpu_cores/8)} if it is executing on a machine with more than 16 processors.
This is a conservative limit that achieves benefits due to parallelism but does not consume all system resources. 
It is not meant to be optimal, and as shown in our evaluation, different scripts achieve optimal speedup with different \ttt{--width} values, which indicates an interesting direction for future work.

\subsection{Backend: From DFGs to a Parallel Shell Script}
\label{backend}

After %
  applying transformations~\sx{ir:transformations}, \sys translates all DFGs back into a shell script.
Nodes of the graph are instantiated with the commands and flags they represent, and edges are instantiated as named pipes.
A prologue in the script creates the necessary intermediate pipes, and a \ttt{trap} call takes care of cleaning up when the script aborts.

\section{Runtime}
\label{impl}
\label{optimizer}

This section describes technical challenges related to the execution of the resulting script and how they are addressed by \sys's custom runtime primitives.

\begin{figure}[t]
\centering
\includegraphics[width=0.42\textwidth]{\detokenize{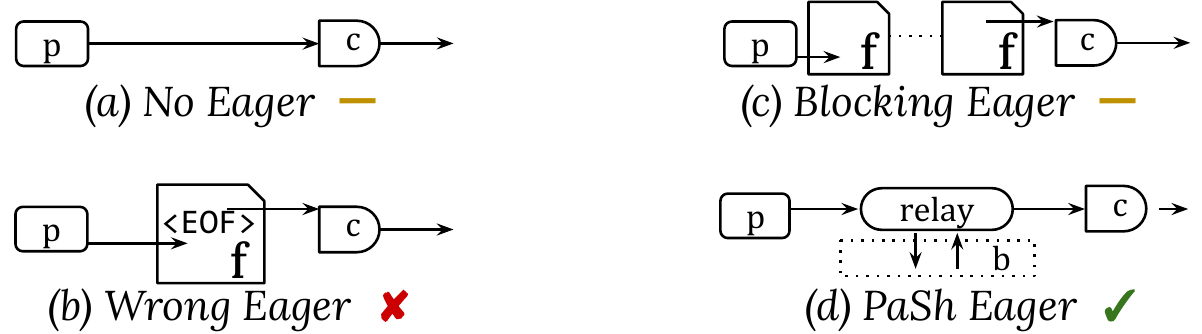}}
\caption{
  \textbf{Eager primitive.}
  Addressing intermediary laziness is challenging:
    (a) FIFOs are blocking;
    (b) files alone introduce race conditions between producer/consumer;
    (c) files + \ttiny{wait} inhibit task parallelism.
Eager \ttiny{relay} nodes (d) address the challenge while remaining within the \sys model.
}
\vspace{-15pt}
\label{fig:eager}
\end{figure}

\heading{Overcoming Laziness}
The shell's evaluation strategy is unusually lazy, in that most commands and shell constructs consume their inputs only when they are ready to process more.
Such laziness leads to CPU underutilization, as commands are often blocked when their consumers are not requesting any input.
Consider the following fragment:

\begin{minted}[fontsize=\footnotesize]{bash}
    mkfifo t1 t2
    grep "foo" f1 > t1 & grep "foo" f2 > t2 & cat t1 t2
\end{minted}

\noindent
The \ttt{cat} command will consume input from \ttt{t2} only after it completes reading from \ttt{t1}.
As a result, the second \ttt{grep} will remain blocked until the first \ttt{grep} completes (Fig.~\ref{fig:eager}a). %

To solve this, one might be tempted to replace FIFOs with files, a central \unix abstraction, simulating pipes of arbitrary buffering (Fig.~\ref{fig:eager}b).
Aside from severe performance implications, naive replacement can lead to subtle race conditions, as a consumer may reach {\sc EOF} before a producer.
Alternatively, consumers could wait for producers to complete before opening the file for reading (Fig.~\ref{fig:eager}c);
  however, this would insert artificial barriers impeding task-based parallelism and wasting disk resources---that is, this approach allows for data parallelism to the detriment of task parallelism.

To address this challenge,
  \sys inserts and instantiates eager \ttt{relay} nodes at these points (Fig.~\ref{fig:eager}d). %
These nodes feature tight multi-threaded loops that consume input eagerly while attempting to push data to the output stream, forcing upstream nodes to produce output when possible while also preserving task-based parallelism.
In \sys's evaluation~\sx{eval}, these primitives have the names presented in Fig.~\ref{fig:eager}.

\heading{Splitting Challenges}
To offer data parallelism, \sys needs to split an input data stream to multiple chunks operated upon in parallel.
Such splitting is needed at least once at the beginning of a parallel fragment, and possibly every time within the parallel program when an \emph{aggregate} function of a stage merges data into a single stream.

To achieve this, \sys's transformations insert split nodes that correspond to a custom \ttt{split} command.
For \ttt{split} to be effective, it needs to disperse its input uniformly. %
\sys does not do this in a round-robin fashion, as that would require augmenting the data stream with additional metadata to maintain FIFO ordering---a challenge for both performance and correctness.
\sys instead splits chunks in-order, which necessitates knowledge of the input size beforehand and which is not always available.
To address this challenge, \sys provides a \ttt{split} implementation that first consumes its complete input, counts its lines, and then splits it uniformly across the desired number of outputs. 
\sys also inserts eager \ttt{relay} nodes after all \ttt{split} outputs (except for the last one) to address laziness as described above.

\begin{table*}[t]
\center
\footnotesize
\caption{
  \footnotesize{
    \textbf{Summary of \unix one-liners}.
    Structure summarizes the different classes of commands used in the script.
    Input and seq. time report on the input size fed to the script and the timing of its sequential execution.
    Nodes and compile time report on \sys's resulting DFG size (which is equal to the number of resulting processes and includes \ttiny{agg}regators, \ttiny{eager}, and \ttiny{split} nodes) and compilation time for two indicative \ttiny{--width}s.
  }
}
\begin{tabular*}{\textwidth}{l @{\extracolsep{\fill}} llllllll}
\toprule
Script ~&~ Structure & Input &Seq. Time & \multicolumn{2}{l}{\#Nodes(16, 64)} &\multicolumn{2}{l}{Compile Time (16, 64)} & Highlights \\
\midrule
nfa-regex ~&~ $3\times\tsta$ & 1~GB & 79m35.197s & 49 & 193 & 0.056s & 0.523s & complex NFA regex \\
sort ~&~ $\tsta, \tpur$ & 10~GB & 21m46.807s & 77 & 317 & 0.090s & 1.083s & \tti{sort}ing \\
top-n ~&~ $2\times\tsta, 4\times\tpur$ & 10~GB & 78m45.872s & 96 & 384 & 0.145s & 1.790s & double \tti{sort}, \tti{uniq} reduction \\
wf ~&~ $3\times\tsta, 3\times\tpur$ & 10~GB & 22m30.048s & 96 & 384 & 0.147s & 1.809s & double \tti{sort}, \tti{uniq} reduction \\
spell ~&~ $4\times\tsta, 3\times\tpur$ & 3~GB & 25m7.560s & 193 & 769 & 0.335s & 4.560s & comparisons (\tti{comm}) \\
difference ~&~ $2\times\tsta, 2\times\tpur, \tnpu$ & 10~GB & 25m49.097s & 125 & 509 & 0.186s & 2.341s & non-parallelizable \tti{diff}ing \\
bi-grams ~&~ $3\times\tsta, 3\times\tpur$ & 3~GB & 38m9.922s & 185 & 761 & 0.313s & 4.310s & stream shifting and merging \\
set-difference ~&~ $5\times\tsta, 2\times\tpur, \tnpu$ & 10~GB & 51m32.313s & 155 & 635 & 0.316s & 4.358s & two pipelines merging to a \tti{comm} \\
sort-sort ~&~ $\tsta, 2\times\tpur$ & 10~GB & 31m26.147s & 154 & 634 & 0.293s & 3.255s & parallelizable \tpur after \tpur \\
shortest-scripts ~&~ $5\times\tsta, 2\times\tpur$ & 85~MB & 28m45.900s & 142 & 574 & 0.328s & 4.657s & long \tsta pipeline ending with \tpur \\
\bottomrule
\end{tabular*}
\label{tab:eval}
\end{table*}

\heading{Dangling FIFOs and Zombie Producers}
Under normal operation, a command exits after it has produced and sent all its results to its output channel.
If the channel is a pipe and its reader exits early, the command is notified to stop writing early.
In \unix, this notification is achieved by an out-of-band error mechanism: the
operating system delivers a \ttt{PIPE} signal to the producer,
notifying it that the pipe's consumer has exited.  This handling is different
from the error handling for other system calls
and unusual compared to non-\unix systems\footnote{For example,
  Windows indicates errors for \ttiny{WriteFile} using its return
  code---similar to \ttiny{DeleteFile} and other Win32 functions.}
primarily because pipes and pipelines are at the heart of \unix.
Unfortunately though, if a pipe has not been opened for writing yet,
  \unix cannot signal this condition.
Consider the following script:
\begin{minted}[fontsize=\footnotesize]{bash}
  mkfifo fifo1 fifo2
  cat in1 > fifo1 & cat in2 > fifo2 &
  cat fifo1 fifo2 | head -n 1 & wait
\end{minted}
\noindent
In the code above, \ttt{head} exits early causing the last \ttt{cat} to exit before opening \ttt{fifo2}.
As a result, the second \ttt{cat} never receives a \ttt{PIPE} signal that its consumer exited---after all, \ttt{fifo2} never even had a consumer!
This, in turn, leaves the second \ttt{cat} unable to make progress, as it is both blocked and unaware of its consumer exiting.
Coupled with \ttt{wait} at the end, the entire snippet reaches a deadlock.

This problem is not unique to \sys;
  it occurs even when manually parallelizing scripts using FIFOs (but not when using \eg intermediary files, \emph{Cf.} \S\ref{impl}, Laziness).
It is exacerbated, however, by \sys's use of the \ttt{cat fifo1 fifo2} construct, used pervasively when parallelizing commands in \sta.

To solve this problem, \sys emits cleanup logic that operates from the
end of the pipeline and towards its start.  The emitted code first
gathers the IDs of the output processes and passes them as parameters
to \ttt{wait}; this causes \ttt{wait} to block only on the output
producers of the dataflow graph.  Right after \ttt{wait}, \sys inserts
a routine that delivers \ttt{PIPE} signals to any remaining processes
upstream.

\heading{Aggregator Implementations}
As discussed earlier, commands in \tpur can be parallelized using a \emph{map} and an \emph{aggregate} stage~\sx{parallelizability}.
\sys implements \emph{aggregate} for several commands in \tpur to enable parallelization.
A few interesting examples are \emph{aggregate} functions for
  (i) \ttt{sort}, which amounts to the merge phase of a merge-sort (and on GNU systems is implemented as \ttt{sort -m}),
  (ii) \ttt{uniq} and \ttt{uniq -c}, which need to check conditions at the boundary of their input streams,
  (iii) \ttt{tac}, which consumes stream descriptors in reverse order, and 
  (iv) \ttt{wc}, which adds inputs with an arbitrary number of elements (\eg \ttt{wc -lw} or \ttt{wc -lwc} \etc).
The \emph{aggregate} functions iterate over the provided stream descriptors, \ie they work with more than two inputs, and apply pure functions at the boundaries of input streams (with the exception of \ttt{sort} that has to interleave inputs).

\section{Evaluation}
\label{eval}

\begin{figure}[t]
\centering
\includegraphics[width=0.37\textwidth]{\detokenize{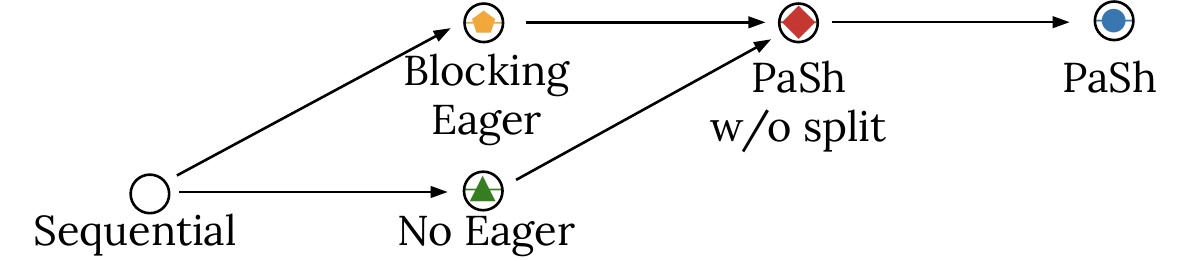}}
\caption{
  \textbf{Runtime setup lattice.}
  Parallel \noeager and \blockingeager improve over sequential, but are not directly comparable.
  \nosplit adds \sys's optimized \ttiny{eager} relay, and \pash uses all primitives in \S\ref{optimizer} (Fig.~\ref{fig:microbenchmarks}).
}
\vspace{-15pt}
\label{fig:lattice}
\end{figure}

This section reports on whether \sys can indeed offer performance benefits automatically and correctly using several scripts collected out from the wild along with a few micro-benchmarks for targeted comparisons.

\heading{Highlights}
This paragraph highlights results for \ttt{width=16}, but \sys's evaluation reports on varying \ttt{width}s (2--64).
Overall, applying \sys to all 44 unmodified scripts accelerates 39 of them by 1.92--17.42$\times$; for the rest, the parallel performance is comparable to the sequential (0.89, 0.91, 0.94, 0.99, 1.01$\times$). The total average speedup over all 44 benchmarks is $6.7\times$.
\sys's runtime primitives offer significant benefits---for the 10 scripts that we measured with and without the runtime primitives they bump the average speedup from $5.9\times$ to $8.6\times$.
\sys significantly outperforms \ttt{sort --parallel}, a hand-tuned parallel implementation, and performs better than GNU \ttt{parallel}, which returns incorrect results if used without care. %

Using \sys's standard library of annotations for POSIX and GNU commands~\sx{parallelizability}, the vast majority of programs ($>40$, with $>200$ commands) require no effort to parallelize other than invoking \sys;
  only 6 ($<3\%$) commands, outside this library, needed a single-record annotation~\sx{macro2}.

In terms of correctness, \sys's results on multi-GB inputs are identical to the sequential ones.
Scripts feature ample opportunities for breaking semantics~\sx{micro}, which \sys avoids.

\begin{figure*}[t]
    \centering
    \includegraphics[width=\textwidth]{\detokenize{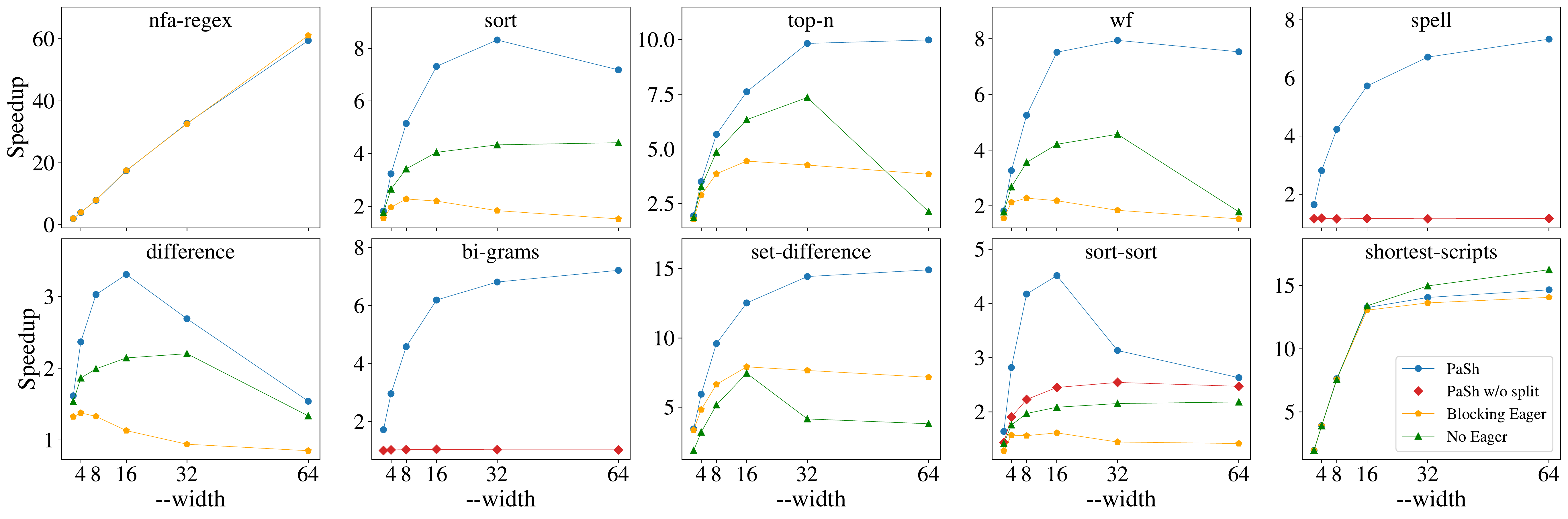}}
    \vspace{-20pt}
    \caption{
      \textbf{\sys's speedup for \ttiny{width=}2--64.}
      Different configurations per benchmark:
      (i) PaSh: the complete implementation with \ttiny{eager} and \ttiny{split} enabled,
      (ii) PaSh w/o split: \ttiny{eager} enabled (no \ttiny{split}),
      (iii) Blocking Eager: only blocking \ttiny{eager} enabled (no \ttiny{split}),
      (iv) No Eager: both \ttiny{eager} and \ttiny{split} disabled.
  For some pairs of configurations, \sys produces identical parallel scripts and thus only one is shown.
}
    \vspace{-12pt}
    \label{fig:microbenchmarks}
\end{figure*}

\heading{Setup}
\sys was run on 512GB of memory and 64 physical $\times$ 2.1GHz Intel Xeon E5-2683 cores, Debian 4.9.144-3.1, GNU Coreutils 8.30-3, GNU Bash 5.0.3(1), and Python 3.7.3---without any special configuration in hardware or software.
Except as otherwise noted,
  (i) all pipelines are set to (initially) read from and (finally) write to the file-system,
  (ii) \ttt{curl} fetches data from a different physical host on the same network connected by 1Gbps links.

We note a few characteristics of our setup that minimize statistical non-determinism:
  (1) our evaluation experiments take several hours to complete (about 23 hours for the full set),
  (2) our experimental infrastructure is hosted on premises, not shared with other groups or researchers, 
  (3) the runtime does not include any managed runtimes, virtualization, or containment,\footnote{
  While \sys is available via Docker too, all results reported in this paper are from non-containerized executions.
}
  (4) many commands are repeated many times---for example, there are more than 40 instances of \ttt{grep} in our benchmark set.
The set of benchmarks also executes with smaller inputs multiple times a week (using continuous integration), reporting minimal statistical differences between runs.

\heading{Parallelism}
\sys's degree of parallelism is configured by the \ttt{--width} flag~\sx{ir:transformations}. 
\sys does not control a script's initial parallelism (\eg a command could spawn 10 processes), and thus the resulting scripts often reach maximum parallelization benefits with a value of \ttt{width} smaller than the physical cores available in our setup (in our case 64).

\subsection{Common \unix One-liners}
\label{ours}

We first evaluate \sys on a set of popular, common, and classic \unix pipeline patterns~\cite{bentley1985spelling, bentley1986literate, taylor2004wicked}.
The goal is to evaluate performance benefits due to \sys's  
  (i) DFG transformations alone, including how \ttt{--width} affects speedup, and
  (ii) runtime primitives, showing results for all points on the runtime configuration lattice (Fig.~\ref{fig:lattice}).

\heading{Programs}
Tab.~\ref{tab:eval} summarizes the first collection of programs.
NFA-Regex is centered around an expensive NFA-based backtracking expression and all of its commands are in \sta.
Sort is a short script centered around a \pur command.
Wf and Top-n are based on McIlroy's classic word-counting program~\cite{bentley1986literate};
  they use sorting, rather than tabulation, to identify high-frequency terms in a corpus.
Spell, based on the original \ttt{spell} developed by Johnson~\cite{bentley1985spelling}, is another \unix classic:
  after some preprocessing, it makes clever use of \ttt{comm} to report words not in a dictionary.
Shortest-scripts extracts the 15 shortest scripts in the user's \ttt{PATH}, using the \ttt{file} utility and a higher-order \ttt{wc} via \ttt{xargs}~\cite[pg. 7]{taylor2004wicked}.
Diff and Set-diff compare streams via a \ttt{diff} (in \npu, non-parallelizable) and \ttt{comm} (in \pur), respectively.
Sort-sort uses consecutive \pur commands without interleaving them with commands that condense their input size (\eg \ttt{uniq}).
Finally, Bi-grams replicates and shifts a stream by one entry to calculate bigrams.

\heading{Results}
Fig.~\ref{fig:microbenchmarks} presents \sys's speedup as a function of \ttt{width=}2--64.
Average speedups of the optimized \sys, \ie with \ttt{eager} and \ttt{split} enabled, for \ttt{width=}\{2, 4, 8, 16, 32, 64\} are \{1.97, 3.5, 5.78, 8.83, 10.96, 13.47\}$\times$, respectively.
For \noeager, \ie \sys' transformations without its runtime support, speedups drop to 1.63, 2.54, 3.86, 5.93, 7.46, 9.35$\times$.

Plots do not include lines for configurations that lead to identical parallel programs.
There are two types of such cases. %
In the first, the \pash (blue) and \nosplit (red, hidden) lines are identical for scripts where \sys does not add \ttt{split}, as the width of the DFG is constant;
  conversely, when both lines are shown (\eg Spell, Bi-grams, and Sort), \sys has added \ttt{split}s due to changes in the DFG width (e.g. due to a \npu command).
In the second type, \emph{Pash w/o Split} (red) is identical to \noeager (green, hidden) and \blockingeager (orange, hidden) because the input script features a command in \pur or \npu relatively early.
This command requires an aggregator, whose output is of width 1, beyond which \nosplit configurations are sequential and thus see no speedup.
Finally, Tab.~\ref{tab:eval} shows that \sys's transformation time is negligible, and its COST~\cite{mcsherryscalability}, \ie the degree of parallelism threshold over which \sys starts providing absolute execution time benefits, is 2. %

\heading{Discussion}
As expected, scripts with commands only in \sta see linear speedup.
\sys's \ttt{split} benefits scripts with \tpur or \tnpu commands, without negative effects on the rest.
\sys's \ttt{eager} primitive improves over \noeager and \blockingeager for all scripts.
\noeager is usually faster than \blockingeager	
since it allows its producer and consumer to execute in parallel.
Sort-sort illustrates the full spectrum of primitives:
  (i) \nosplit offers benefits despite the lack of \ttt{split} because it fully parallelizes the first \ttt{sort}, and 
  (ii) \pash gets full benefits because \ttt{split}ting allows parallelizing the second \ttt{sort} too.

As described earlier, \sys often achieves the maximum possible speedup %
  for a \ttt{width} that is lower than the number of available cores---\ie \ttt{width=}16--32 for a 64-core system.
This is also because \sys's runtime primitives spawn new processes---\eg Sort with \ttt{width=}8 spawns 37 processes:
  8 \ttt{tr}, 8 \ttt{sort}, 7 \emph{aggregate}, and 14 \ttt{eager} processes.

\heading{Take-aways}
\sys accelerates scripts by up to 60$\times$, depending on the characteristics of the commands involved in a script.
Its runtime constructs improve over the baseline speedup achieved by its parallelization transformations.

\subsection{Unix50 from Bell Labs}
\label{unix50}

We now turn to a set of \unix pipelines found out in the wild.

\heading{Programs}
In a recent celebration of \unix's 50-year legacy, Bell Labs created 37 challenges~\cite{unix50} solvable by \unix pipelines.
The problems were designed to highlight \unix's modular philosophy~\cite{mcilroy1978unix}.
We found unofficial solutions to all-but-three problems on GitHub~\cite{unix50sol}, expressed as pipelines with 2--12 stages (avg.: 5.58).
They make extensive use of standard commands under a variety of flags, and appear to be written by non-experts (contrary to \S\ref{ours}, they often use sub-optimal or non-\unix-y constructs).
\sys executes each pipeline as-is, without any modification.

\begin{figure}[t]
  \centering
  \includegraphics[width=\columnwidth]{\detokenize{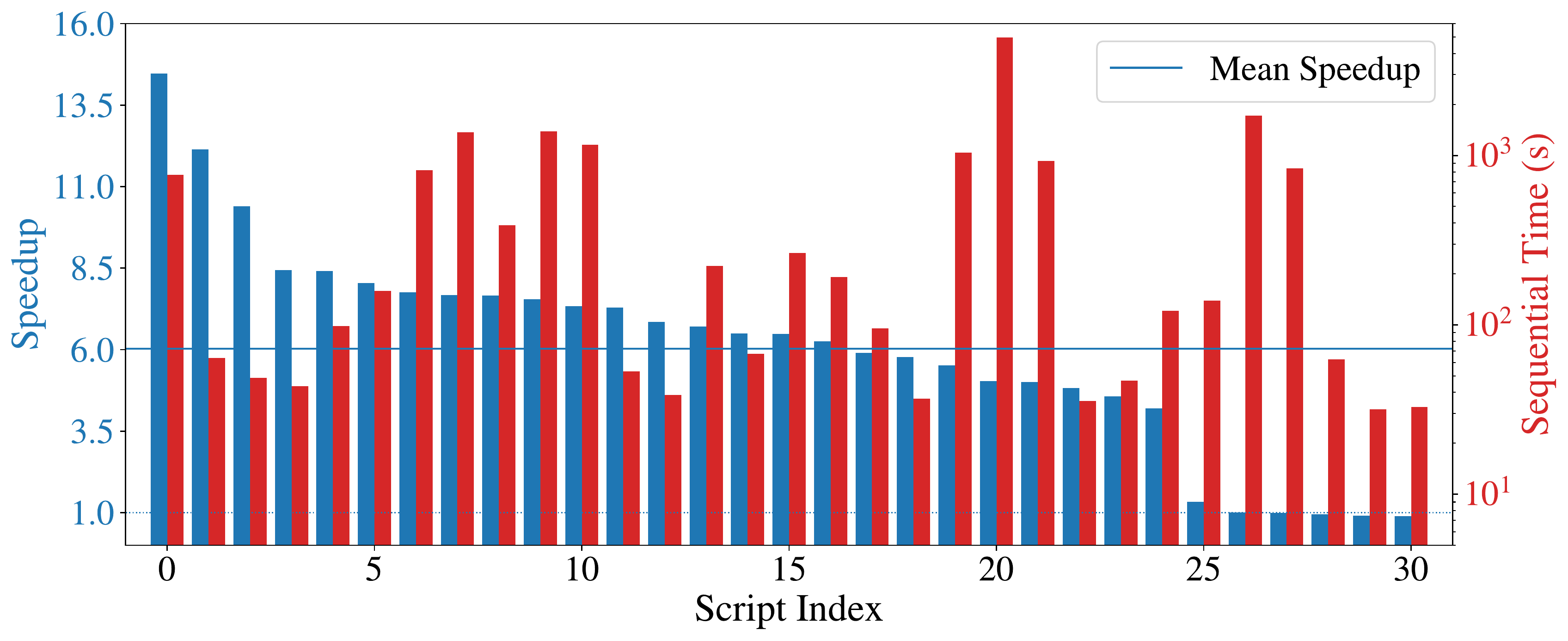}}
  \vspace{-15pt}
  \caption{
	  \textbf{Unix50 scripts.}
    Speedup (left axis) over sequential execution (right axis) for Unix50 scripts.
    Parallelism is 16$\times$ on 10GB of input data~\cf{unix50}.
    Pipelines are sorted in descending speedup order. %
  }
  \vspace{-15pt}
  \label{fig:unix50-individual}
\end{figure}

\heading{Results}
Fig.~\ref{fig:unix50-individual} shows the speedup (left) over the sequential runtime (right) for 31 pipelines, with \ttt{width=}16 and 10GB inputs.
It does not include 3 pipelines that use \ttt{head} fairly early thereby finishing execution in under $0.1$ seconds.
We refer to each pipeline using its x-axis index (\#0--30) in Fig.~\ref{fig:unix50-individual}.
Average speedup is $6.02\times$, %
  and weighted average (with the absolute times as weights) is $5.75\times$.

\heading{Discussion}
Most pipelines see significant speedup, except \#25-30
  that see no speedup because they contain general commands that \sys cannot parallelize without risking breakage---\eg \ttt{awk} and \ttt{sed} \ttt{-d}. %
A \unix expert would notice that some of them can be replaced with \unix-specific commands---\eg \ttt{awk "{print \$2, \$0}" | sort -nr}, used to sort on the second field can be replaced with a single \ttt{sort -nr -k 2} (\#26).
The targeted expressiveness of the replacement commands can be exploited by \sys---in this specific case, achieving 8.1$\times$ speedup (\vs the original 1.01$\times$).

For all other scripts (\#0--24), \sys's speedup is capped due to a combination of reasons:
  (i) scripts contain pure commands that are parallelizable but don't scale linearly, such as \ttt{sort} 
  (\#5, 6, 7, 8, 9, 19, 20, 21, 23, 24), 
  (ii) scripts are deep pipelines that already exploit task parallelism
  (\#4, 10, 11, 13, 15, 17, 19, 21, 22),
or 
  (iii) scripts are not CPU-intensive, resulting in pronounced I/O and constant costs 
  (\#3, 4, 11, 12, 14, 16, 17, 18, 22).

\heading{Take-aways} \sys accelerates unmodified pipelines found in the wild;
  small tweaks can yield further improvements, showing that \sys-awareness and scripting expertise can improve performance.
Furthermore, \sys does not significantly decelerate non-parallelizable scripts.

\subsection{Use Case: NOAA Weather Analysis}
\label{macro1}

We now turn our attention to Fig.~\ref{fig:example}'s script~\sx{bg}.

\heading{Program}
This program is inspired by the central example in ``Hadoop: The Definitive Guide''~\cite[\S2]{hadoop:15}, where it exemplifies a realistic analytics pipeline comprising 3 stages:
  fetch NOAA data (shell), convert them to a Hadoop-friendly format (shell), and calculate the maximum temperature (Hadoop).
While the book focuses only on the last stage, \sys parallelizes the entire pipeline.

\heading{Results}
The complete pipeline executes in 44m2s for five years (82GB) of data.
\sys with \ttt{width=}16 leads to 2.52$\times$ speedup, with different phases seeing different benefits:
  2.04$\times$ speedup (vs. 33m58s) for all the pre-processing ($75\%$ of the total running time) and 
  12.31$\times$ speedup (vs. 10m4s) for computing the maximum.

\heading{Discussion}
The speedup of the preprocessing phase of the pipeline is bound by the network and I/O costs since \ttt{curl} downloads 82GB of data.
However, the speedup for the processing phase (CPU-bound) is $12.31\times$, much higher than what would be achieved by parallelizing per year (for a total of five years).
Similar to Unix50~\sx{unix50}, we found that large pipelines enable significant freedom in terms of expressiveness.

\heading{Take-aways}
\sys can be applied to programs of notable size and complexity to offer significant acceleration.
\sys is also able to extract parallelism from fragments that are not purely compute-intensive, \ie the usual focus of conventional parallelization systems.

\subsection{Use Case: Wikipedia Web Indexing}
\label{macro2}

We now apply \sys to a large web-indexing script. %

\heading{Program}
This script reads a file containing Wikipedia URLs, downloads the pages,
extracts the text from HTML, and applies natural-language processing---\eg trigrams, character conversion, term frequencies---to index it.
It totals 34 commands written in multiple programming languages.

\heading{Results}
The original script takes 191min to execute on 1\% of Wikipedia (1.3GB).
With \ttt{width=}16, \sys brings it down to 15min ($12.7\times$), with the majority of the speedup coming from the HTML-to-text conversion. %

\heading{Discussion}
The original script contains 34 pipeline stages, thus the sequential version already benefits from task-based parallelism.
It also uses several utilities not part of the standard POSIX/GNU set---\eg its \ttt{url-extract}ion is written in JavaScript and its \ttt{word-stem}ming is in Python.
\sys can still operate on them as their parallelizability properties---\sta for \ttt{url-extract} and \ttt{word-stem}---can be trivially described by annotations.
Several other stages are in \sta allowing \sys to achieve benefits by exposing data parallelism.

\heading{Take-aways}
\sys operates on programs with (annotated) commands outside the POSIX/GNU subsets and leads to notable speedups, even when the original program features significant task-based parallelism.

\subsection{Further Micro-benchmarks}
\label{micro}

As there are no prior systems directly comparable to \sys, we now draw comparisons with two specialized cases that excel within smaller fragments of \sys's proposed domain.

\pichskip{15pt}%
\parpic[r][t]{%
  \begin{minipage}{50mm}
    \includegraphics[width=\linewidth]{\detokenize{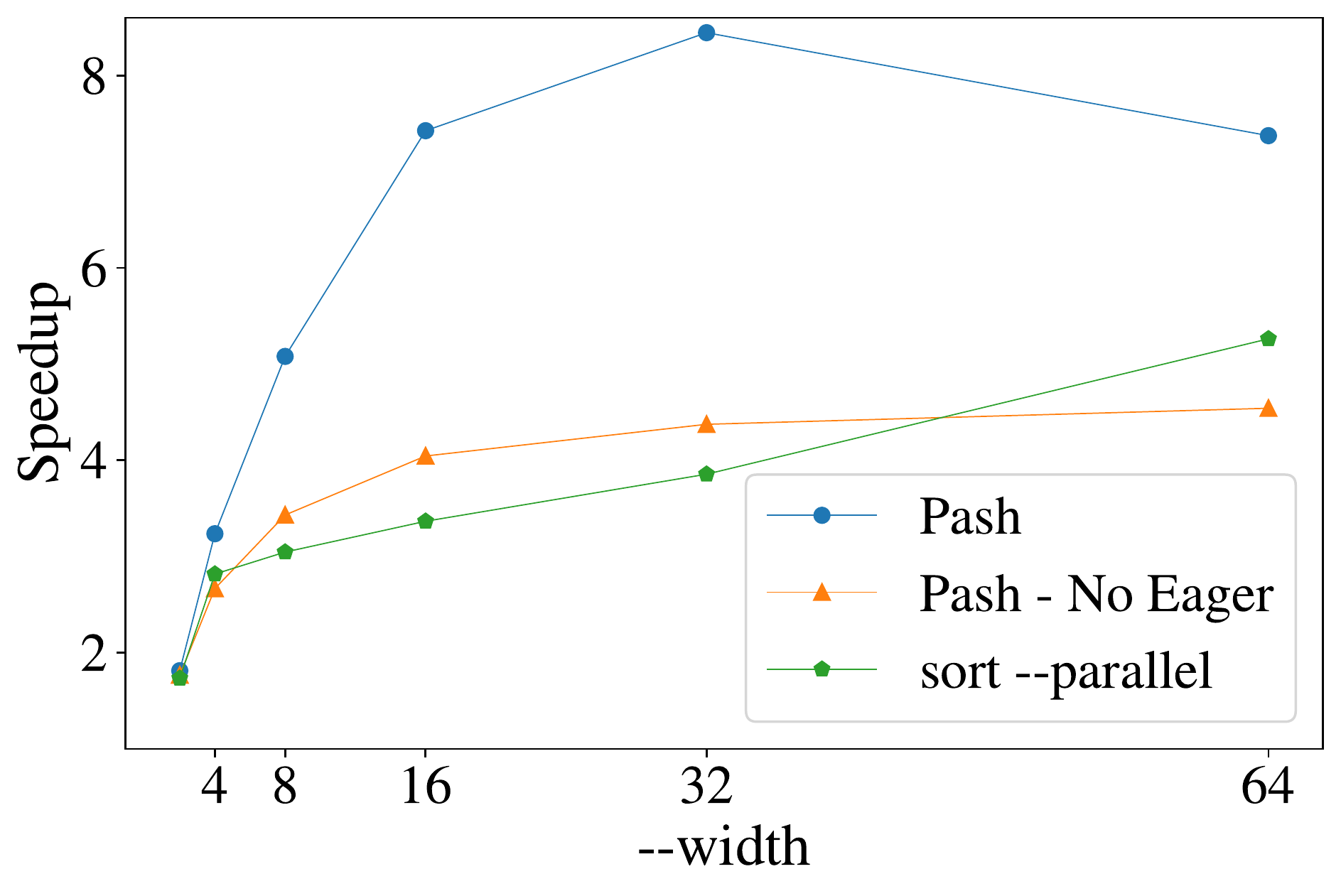}}%
  \end{minipage}
}
\heading{Parallel Sort}
First, we compare a \ttt{sort} parallelized by \sys ($S_{p}$) against the same \ttt{sort} invoked using the 
\ttt{--parallel} flag set ($S_{g}$).\footnote{
  Both \ttiny{sort}s use the same buffer size internally~\cite{sortp:15}.
}
While the \ttt{--parallel} flag is not a general solution, the comparison serves to establish a baseline for \sys.
$S_{g}$'s parallelism is configured to $2\times$ that of $S_{p}$'s \ttt{--width} (\ie the rightmost plot point for $S_{g}$ is for \ttt{--parallelism=128}), 
to account for \sys's additional runtime processes.

A few points are worth noting.
$S_{p}$ without \ttt{eager} performs comparably to $S_{g}$, and with \ttt{eager} it outperforms $S_{g}$ ($\sim2\times$);
  this is because \ttt{eager} adds intermediate buffers %
  that ensure CPU utilization is high.
$S_{g}$ indicates that \ttt{sort}'s scalability is inherently limited (\ie due to \ttt{sort}, not \sys);
  this is why all scripts that contain \ttt{sort} (\eg \S\ref{ours}--\ref{macro2}) are capped at $8\times$ speedup.
The comparison also shows \sys's benefits to command developers:
  a low-effort parallelizability annotation achieves better scalability than a custom flag (and underlying parallel implementation) manually added by developers.

\heading{GNU Parallel}
We compare \sys to \ttt{parallel} (v.20160422), a GNU utility for running other commands in parallel~\cite{Tange2011a}, on a small bio-informatics script.
Sequential execution takes 554.8s \vs \sys's 128.5s (4.3$\times$), with most of the overhead coming from a single command---\ttt{cutadapt}.

There are a few possible ways users might attempt to use GNU \ttt{parallel} on this program.
They could use it on the bottleneck stage, assuming they can deduce it, bringing execution down to 304.4s ($1.8\times$ speedup).
Alternatively, they could (incorrectly) sprinkle \ttt{parallel} across the entire program.
This would lead to 3.2$\times$ performance improvements but incorrect results with respect to the sequential execution---with 92\% of the output showing a difference between sequential and parallel execution.
\sys's conservative program transformations are not applied in program fragments with unclear parallelizability properties.

\section{Related Work}
\label{related}

Existing techniques for exploiting parallelism are not directly comparable to \sys, because they either require significantly more \emph{user} effort (see \S\ref{intro} for the distinction between users and developers) or are too specialized, targeting narrow domains or custom programming abstractions.

\heading{Parallel Shell Scripting}
Utilities exposing parallelism on modern \unix{}es---\eg \ttt{qsub}~\cite{gentzsch2001sun}, \textsc{SLURM}~\cite{yoo2003slurm}, %
  \ttt{parallel}~\cite{Tange2011a}---are limited to embarrassingly parallel (and short) programs and are predicated upon explicit and careful user invocation:
  users have to navigate through a vast array of different configurations, flags, and modes of invocation to achieve parallelization without jeopardizing correctness. %
For example, \ttt{parallel} contains flags such as \ttt{--skip-first-line}, \ttt{-trim}, and \ttt{--xargs},
and introduces (and depends on) other programs with complex semantics, such as ones for SQL querying and CSV parsing.
In contrast, \sys manages to parallelize large scripts correctly with minimal-to-zero user effort.

Several shells~\cite{duff1990rc, mcdonald1988support, dagsh:17} add primitives for non-linear pipe topologies---some of which target parallelism.
Here too, however, users are expected to manually rewrite scripts to exploit these new primitives, contrary to \sys.

Recently, Greenberg~\cite{smoosh:18} argued that the shell and its constructs can be seen as a DSL for orchestrating concurrent processes.
\sys's extraction of dataflow regions is based on a similar observation, but its central focus is on achieving data parallelism from these dataflow regions automatically.

Developed 
concurrently 
with \sys, the Process-Offload SHell (POSH)~\cite{raghavan2020posh} is a shell and runtime that automatically reduces data movement when running shell pipelines on data stored in remote storage \'a la NFS.
POSH accelerates I/O-heavy pipelines that access files in remote filesystems, by offloading computation to servers closer to the data.
\sys is a shell-to-shell compiler that parallelizes \unix shell scripts running on a single multi-processor machine by transforming them to DFGs, applying transformations, and then transforming them back to parallel shell scripts augmented with \sys's runtime primitives that are executed on the user's shell. 
Both PaSh and POSH observe that \unix commands can have arbitrary behaviors~\sx{bg:challenges}, thus each introducing an annotation language that fits its problem:
  POSH uses annotations to identify which files are accessed by a pipeline, and thus co-locates commands and their dependencies;
  PaSh uses annotations to identify whether a command is parallelizable and, if so, how to translate it to a dataflow node.
Both systems descend from a lineage of annotation-based black-box transformations~\cite{ignis:19, mozart:19, breakapp:ndss:2018, oa}.

\heading{Low-level Parallelization}
There exists significant work on automating parallelization at the instruction level, starting with explicit \ttt{DOALL} and \ttt{DOACROSS} annotations~\cite{par1, par2} and continuing with compilers that attempt to automatically extract parallelism~\cite{padua1993polaris,hall1996maximizing}.
These efforts operate at a lower level than \sys (\eg that of instructions or loops rather than the boundaries of programs that are part of a script), within a single-language or single-target environments, and require source modifications.

More recent work focuses on extracting parallelism from domain-specific programming models~\cite{cilk5, streamIt, galois} and interactive parallelization tools~\cite{parascope, ipat}.
These tools simplify the expression of parallelism, but still require significant user involvement in discovering and exposing parallelism.

\heading{Correct Parallelization of Dataflow Graphs}
The DFG is a prevalent model in several areas of data processing, including batch-~\cite{mapreduce:08, spark:12} and stream-processing ~\cite{murray2013naiad, carbone2015flink}.
Systems implementing DFGs often perform optimizations that are correct given subtle assumptions on the dataflow nodes that do not always hold, introducing erroneous behaviors.
Recent work~\cite{HSSGG2014, SHGW2015, MSAIT2019, kallas2020diffstream} attempts to address this issue by performing optimizations only in cases where correctness is preserved, or by testing that applied optimizations preserve the original behavior.
\sys draws inspiration from these efforts, in that it delegates the satisfaction of assumptions to the annotation writers, who are expected to be command developers rather than shell users~\sx{intro}, ensuring that transformations preserve the behavior of the original dataflow.
Its  DFG model, however, is different from earlier efforts in that it explicitly captures and manipulates ordering constraints.
The constraints are due to the intricacies of the \unix model---\eg FIFO streams, argument processing, and concatenation operators.

\heading{Parallel Userspace Environments}
By focusing on simplifying the development of distributed programs, a plethora of environments additionally assist in the construction of parallel software.
Such systems~\cite{ousterhout1988sprite, mullender1990amoeba, barak1998mosix}, languages~\cite{erlang:96, acute:05, mace:07}, or system-language hybrids~\cite{pike1990plan9, andromeda:15, cloudhaskell:11} hide many of the challenges of dealing with concurrency as long as developers leverage the provided abstractions---which are strongly coupled to the underlying operating or runtime system.
Even shell-oriented efforts such as Plan9's \ttt{rc} are not backward-compatible with the \unix shell, and often focus primarily on hiding the existence of a network rather than automating parallel processing.

\heading{Parallel Frameworks}
Several frameworks~\cite{streamit:02, brook:04, phoenix:11, raftlib:17, fetterly2009dryadlinq} offer fully automated parallelism as long as special primitives are used---\eg map-reduce-style primitives for Phoenix~\cite{phoenix:11}.
These primitives make strong assumptions about the nature of the computation---\eg commutative and associative aggregation functions that can be applied on their inputs in any order.
By targeting specific classes of computation (\emph{viz.} \sys's parallelizability), these primitives are significantly optimized for their target domains.
\sys instead chooses an approach that is better tailored to the shell:
  it does not require rewriting parts of a shell script using specific parallelization-friendly primitives, but rather lifts arbitrary commands to a parallelization-friendly space using an annotation framework.

Dryad~\cite{isard2007dryad} is a distributed system for dataflow graphs. Dryad offers a scripting language, Nebula, that allows using shell commands such as \ttt{grep} or \ttt{sed} in place of individual dataflow nodes. The main difference with \sys is that in Dryad the programmer needs to explicitly express the dataflow graph, which is then executed in a distributed fashion, whereas \sys automatically parallelizes a given shell script by producing a parallel script that runs on an unmodified shell of choice.

\section{Conclusion}
\label{discussion}

Shell programs are ubiquitous, use blocks written in a plethora of programming languages, and spend a significant fraction of their time interacting with the broader environment to download, extract, and process data---falling outside the focus of conventional parallelization systems.
This paper presents \sys, a system that allows shell users to parallelize shell programs mostly automatically.
\sys can be viewed as 
  (i) a source-to-source compiler that transforms scripts to DFGs, parallelizes them, and transforms them back to scripts, coupled with 
  (ii) a  runtime component that addresses several practical challenges related to performance and correctness.
\sys's extensive evaluation over 44 unmodified Unix scripts demonstrates non-trivial speedups (0.89--61.1$\times$, avg: 6.7$\times$).

\sys's implementation, as well as all the example code and benchmarks presented in this paper, are all open source and available for download:
\href{https://github.com/andromeda/pash}{github.com/andromeda/pash}.

\begin{acks}
We want to thank 
  Andr\'e DeHon,
  Ben Karel,
  Caleb Stanford,
  Thurston Dang,
  Jean-S\'ebastien L\'egar\'e, 
  Nick Roessler,
  Sage Gerard, and several open-source contributors.
We are grateful to our shepherd, Julia Lawall, for her guidance.
This material is based upon work supported by DARPA contract no. HR00112020013 and no. HR001120C0191, and NSF awards CCF 1763514 and 2008096.
Any opinions, findings, conclusions, or recommendations expressed in this material are those of the authors and do not necessarily reflect those of DARPA or NSF.
\end{acks}

{\small
\bibliography{./bib}
}

\appendix

\section{Annotation for the Command \texttt{cut}}
\label{annotation-apx}

The code below shows the full annotation for \ttt{cut}.

\begin{minted}[fontsize=\footnotesize]{json}
{ "command": "cut",
  "cases": [
    { "predicate": {
        "operator": "or",
        "operands": [
          { "operator": "val_opt_eq",
            "operands": [ "-d", "\n" ] },
          { "operator": "exists",
            "operands": [ "-z" ] }
        ]
      },
      "class": "pure",
      "inputs": [ "args[:]" ],
      "outputs": [ "stdout" ]
    },
    { "predicate": "default",
      "class": "stateless",
      "inputs": [ "args[:]" ],
      "outputs": [ "stdout" ]
    }
  ],
  "options": [ "stdin-hyphen", "empty-args-stdin" ],
  "short-long": [
    { "short": "-d", "long": "--delimiter" },
    { "short": "-z", "long": "--zero-terminated" }
  ]
}
\end{minted}

\section{Artifact Appendix}
\label{annotation-apx}

\heading{Summary}
The artifact consists of several parts: 
  (i) a mirror of \sys' GitHub repository (git commit \texttt{e5f56ec}, available permanently in branch \texttt{eurosys-2021-aec-frozen}) including annotations, the parallelizing compiler, and the runtime primitives presented in this paper;
  (ii) instructions for pulling code and experiments, building from source, preparing the environment, and running the experiments;
  (iii) a 20-minute video walk-through of the entire artifact; and
  (iv) instructions for directly pulling a pre-built Docker container and building a Docker image from scratch;
  (v) scripts, descriptions, and instructions to run the experiments (automatically or manually) to reproduce the graphs and results presented in the paper.

\begin{table}[t]
\center
\setlength\tabcolsep{3pt}
\caption{
  \footnotesize{
    \textbf{Major experiments presented in the paper}.
    There are four major experiments presented in the paper: 
      (i) Common \unix one-liners,
      (ii) Unix50 from Bell Labs, 
      (iii) NOAA Weather Analysis, and
      (iv) Wikipedia Web Indexing.
  }
}
\begin{tabular*}{\columnwidth}{l @{\extracolsep{\fill}} lll}
\toprule
Experiment               &  Section                       & Location  \\
\midrule
  Common \unix one-liners  & \S\ref{ours}                 & \href{https://github.com/andromeda/pash/tree/main/evaluation/benchmarks/oneliners}{https://git.io/JYi9m} \\
  Unix50 from Bell Labs    & \S\ref{unix50}               & \href{https://github.com/andromeda/pash/tree/main/evaluation/benchmarks/unix50}{https://git.io/JYi9n}    \\
  NOAA Weather Analysis    & \S\ref{macro1}               & \href{https://github.com/andromeda/pash/tree/main/evaluation/benchmarks/max-temp}{https://git.io/JYi9C}  \\
  Wikipedia Web Indexing   & \S\ref{macro2}               & \href{https://github.com/andromeda/pash/tree/main/evaluation/benchmarks/web-index}{https://git.io/JYi98} \\
\bottomrule
\end{tabular*}
\label{tab:structure}
\end{table}

\heading{Codebase information} 
Below is a summary of key information about \sys's repository:

\begin{itemize}
  \item Repository: \href{https://github.com/andromeda/pash}{https://github.com/andromeda/pash}
  \item License: MIT
  \item Stats: 2,278 commits, from 14 contributors
\end{itemize}

\heading{Artifact requirements}
Below is a summary of requirements for running \sys and its evaluation experiments:

\begin{itemize}
  \item CPU: a modern multi-processor, to show performance results (the more cpus, the merrier)
  \item Disk: about 10GB for small-input (quick) evaluation, about 100GB+ for full evaluation
  \item Software: Python 3.5+, Ocaml 4.05.0, Bash 5+, and GNU Coreutils (details below)
  \item Time: about 30min for small-input, about 24h for full evaluation
\end{itemize}

\heading{Dependencies}
The artifact depends on several packages; on Ubuntu 18.04:
 libtool,
 m4,
 automake,
 opam,
 pkg-config,
 libffi-dev,
 python3,
 python3-pip,
 wamerican-insane,
 bc,
 bsdmainutils, curl, and wget.
\sys and its experimental and plotting infrastructure make use of the following Python packages: jsonpickle, PyYAML, numpy, matplotlib.
Experiments and workloads have their own dependencies---\eg pandoc-2.2.1, nodejs, and npm (Web indexing), or p7zip-full (Wikipedia dataset).

\heading{Access}
\sys is available via several means, including:

\begin{itemize}
  \item Git: \texttt{git clone git@github.com:andromeda/pash.git}
  \item Docker: \texttt{curl img.pash.ndr.md | docker load}
  \item HTTP: \texttt{wget pkg.pash.ndr.md}
  \item Shell: \texttt{curl -s up.pash.ndr.md | sh}
\end{itemize}

\heading{Code Structure}
This repo hosts the core \sys development. The artifact's directory structure is as follows:

\begin{itemize}
\item \href{https://github.com/andromeda/pash/tree/main/annotations}{annotations}: Parallelizability study and associated command annotations.
\item \href{https://github.com/andromeda/pash/tree/main/compiler}{compiler}: Shell-dataflow translations and associated parallelization transformations.
\item \href{https://github.com/andromeda/pash/tree/main/docs}{docs}: Design documents, tutorials, installation instructions, \etc
\item \href{https://github.com/andromeda/pash/tree/main/evaluation}{evaluation}: Shell pipelines and example scripts used in the evaluation of \sys.
\item \href{https://github.com/andromeda/pash/tree/main/runtime}{runtime}: Runtime component---\eg eager, split, and associated aggregators.
\item \href{https://github.com/andromeda/pash/tree/main/scripts}{scripts}: Scripts related to installation, continuous integration, deployment, and testing.
\end{itemize}

\heading{Calling \sys}
To parallelize a script \texttt{hello-world.sh} with a parallelization degree of 2, from the top-level directory of the repository run:
\begin{minted}[fontsize=\small]{bash}
  ./pa.sh hello-world.sh
\end{minted}
\sys will compile and execute \texttt{hello-world.sh} on the fly.

\heading{Tutorial}
To go through a longer tutorial, see \href{https://github.com/andromeda/pash/blob/main/docs/tutorial.md}{docs/tutorial}.

\heading{Available subcommands}
Run \ttt{./pa.sh --help} to get more information about the available \sys subcommands:

\begin{minted}[fontsize=\footnotesize]{text}
Usage: pa.sh [-h] [--preprocess_only] [--output_preprocessed] 
             [-c COMMAND] [-w WIDTH] [--no_optimize] 
             [--dry_run_compiler] [--assert_compiler_success] 
             [-t] [-p] [-d DEBUG] [--log_file LOG_FILE] 
             [--no_eager] [--speculation {no_spec,quick_abort}]
             [--termination {clean_up_graph,drain_stream}]
             [--config_path CONFIG_PATH] [-v] [input]

Positional arguments:
 input                 The script to be compiled and executed.

optional arguments:
 -h, --help            
                       Show this help message and exit.
 --preprocess_only     
                       Pre-process (not execute) input script.
 --output_preprocessed
                       Output the preprocessed script.
 -c COMMAND, --command COMMAND
                       Evaluate the following COMMAND as a 
                       script, rather than a file.
 -w WIDTH, --width WIDTH
                       Set degree of data-parallelism.
 --no_optimize         
                       Not apply transformations over the DFG.
 --dry_run_compiler    
                       Not execute the compiled script, even 
                       if the compiler succeeded.
 --assert_compiler_success
                       Assert that the compiler succeeded 
                       (used to make tests more robust).
 -t, --output_time     
                       Output the time it took for every step.
 -p, --output_optimized
                       Output the parallel script for inspection.
 -d DEBUG, --debug DEBUG
                       Configure debug level; defaults to 0.
 --log_file LOG_FILE   
                       Location of log file; defaults to stderr.
 --no_eager            
                       Disable eager nodes before merging nodes.
 --termination {clean_up_graph,drain_stream}
                       Determine the termination behavior of the 
                       DFG. Defaults to cleanup after the last 
                       process dies, but can drain all streams 
                       until depletion.
 --config_path CONFIG_PATH
                       Determine the config file path, by
                       default 'PASH_TOP/compiler/config.yaml'.
 -v, --version         Show program's version number and exit
\end{minted}

\end{document}